\documentclass[11pt]{article}

\RequirePackage{amsthm,amsmath,amsfonts,amssymb}
\RequirePackage[numbers]{natbib}
\RequirePackage[colorlinks,citecolor=blue,urlcolor=blue,backref=page,backref=page]{hyperref}
\RequirePackage{graphicx}
\RequirePackage{float}
\usepackage{tikz}
\usepackage[ruled,vlined]{algorithm2e}
\usetikzlibrary{3d, calc,decorations.markings}
\usepackage[margin=1in]{geometry}
\usepackage{setspace}
\usepackage{algorithmic}

\newtheorem{proposition}{Proposition}[section]

\theoremstyle{plain}

\newtheorem{theorem}{Theorem}[section]
\newtheorem{lemma}[theorem]{Lemma}
\newtheorem{corollary}{Corollary}[section]

\theoremstyle{definition}
\newtheorem{definition}[theorem]{Definition}

\title{Enhancing the Tensor Normal via Geometrically Parameterized Cholesky Factors}
\author{
    Quinn Simonis and Martin T. Wells \thanks{Department of Statistics and Data Science, Cornell University; qas3@cornell.edu, mtw1@cornell.edu}
}
\date{Cornell University\\April 2025}

\begin{document}
\maketitle

\begin{abstract}
  In this article, we explore Bayesian extensions of the tensor normal model through a geometric expansion of the multi-way covariance's Cholesky factor inspired by the Fr\'echet mean under the log-Cholesky metric. Specifically, within a tensor normal framework, we identify three structural components in the covariance of the vectorized data. By parameterizing vector normal covariances through such a Cholesky factor representation, analogous to a finite average of multiway Cholesky factors, we eliminate one of these structural components without compromising the analytical tractability of the likelihood, in which the multiway covariance is a special case. Furthermore, we demonstrate that a specific class of structured Cholesky factors can be precisely represented under this parameterization, serving as an analogue to the Pitsianis-Van Loan decomposition. We apply this model using Hamiltonian Monte Carlo in a fixed-mean setting for two-way covariance relevancy detection of components, where efficient analytical gradient updates are available, as well as in a seasonally-varying covariance process regime.
\end{abstract}
\noindent
\begin{small}
\end{small}

\section*{Introduction}
This article explores the generalization of a classical likelihood function for multimodal data, where observations are represented as tensors $\mathcal{Y} \in \mathbb{R}^{\times_{i = 1}^{D} d_{i}}$. The normal tensor model assumes Gaussianity in each mode, which means that individual elements follow $y_{i_{1}, i_{2}, \cdots, i_{D}} \sim \mathcal{N}(0,1)$. Mode-wise products enable the construction of a broad class of distributions that capture correlations along specific modes. Alternatively, tensor normal models can be viewed through the lens of vectorization. Assuming lexicographic order, if $\mathcal{Y} \sim \mathcal{T}\mathcal{N}(M, \{\Sigma_{i}\}_{i = 1}^{D})$, then its vectorized form follows $vec(Y) \sim \mathcal{N}(vec(M), \otimes_{i = D}^{1} \Sigma_{i})$.

This likelihood is particularly valuable in Bayesian analysis for examining correlations in high-dimensional data, as it naturally decomposes complex dependencies into lower-dimensional components. Applications of the tensor normal model include analysis of mortality data \cite{fosdick2014separable}, assessment of healthcare \cite{hatfield2018separable}, and modeling in space time \cite{chen2021space, genton2007separable}. Beyond the two-way case, higher-order examples arise in relational data analysis \cite{hoff2011separable}, where a Bayesian approach extends the tensor normal model through the Tucker product.

Although the tensor normal model offers computational advantages, it imposes structural assumptions on global covariance that may not accurately capture the underlying data-generating process \cite{brown2000blur, song2023separability}.

Our work is closely aligned with the literature on Kronecker product expansions of covariances. This area of work was inspired by the work of \cite{pitsanis1997kronecker}, where the authors show that in fact any arbitrary matrix $Q \in \mathbb{R}^{d_{1} d_{2} \times d_{1}d_{2}}$ can be represented as:
\[
Q = \sum_{i = 1}^{r^{2}} A_{i} \otimes B_{i}, \quad A_{i} \in \mathbb{R}^{d_{1} \times d_{1}},\text{ } B_{i} \in \mathbb{R}^{d_{2} \times d_{2}}, \text{ } r = \min \{d_{1}, d_{2} \}.
\]

Literature in Kronecker product expansions of covariances has attracted attention within a statistical context, wherein the goal is often to find a minimal $d$ to adequately represent a covariance matrix as a sum of Kronecker products:
\begin{equation}
    \Sigma \approx \sum_{i = 1}^{d} A_{i} \otimes B_{i}, \quad A_{i} \in \mathcal{P}^{+}(d_{1}), B \in \mathcal{P}^{+}(d_{2}),
\end{equation}
where $\mathcal{P}^{+}(d)$ denotes the manifold of $d\times d$ symmetric positive definite (SPD) matrices. Examples include \cite{tsiligkaridis2013covariance}, who investigated using Kronecker product expansions to model matrix-valued data using a penalized least squares approach. 

In this article, we develop a Bayesian model selection procedure to identify the level of structural sparsity in the strictly lower triangular entries of the Cholesky factor for a multiway covariance. This approach leverages a geometrically inspired parametrization of Cholesky factors for multiway covariances, addressing one of the key structural assumptions imposed by the tensor normal model.

Our work is motivated by the log-Cholesky metric \cite{lin2019riemannian}, which observes that Cholesky factors of symmetric positive definite (SPD) matrices form a topological manifold, wherein the author endows this manifold with a Riemannian metric that has a closed form Fr\'echet mean. Building on this, we express the Cholesky factor of an arbitrary SPD matrix of Cholesky factors of multiway SPD matrices, introducing a simplex-distributed random variable to regulate the structural complexity of lower triangular entries.  This alternative parametrization maintains full analytic expressivity, and we employ Hamiltonian Monte Carlo \cite{neal2011mcmc} to avoid the constraints of conditional conjugacy required by Gibbs sampling.

The remainder of the paper is structured as follows:
\begin{itemize}
    \item Section \ref{sec: Mathematical Background} provides the mathematical background of the paper, with an introduction to Hamiltonian Monte Carlo in Section \ref{subsec: Bayesian Inference}, and the Tensor Normal distribution, and its structural limitations in Section \ref{subsection: Tensor Normal}.
    \item Section \ref{sec: extending the multi-way parameterization} introduces the differential geometry of Cholesky factors in Section \ref{sec: Differential Geometry of Cholesky Factors} used to extend the multiway covariance parameterization of the tensor normal in Section \ref{subsec: Enhancing the Tensor Normal}. The latter subsection also provides bounds on approximation quality of arbitrary Cholesky factors of symmetric positive definite matrices under our parameterization.
    \item Section \ref{sec: The Bayesian Model} provides auxiliary results for centering matrix priors in section \ref{sec: Centering Matrix Priors}, and two Bayesian models are proposed in Sections \ref{sec: Non-Dynamic} and \ref{sec: seasonally dynamic covariance} for static and seasonally dynamic covariance implementations, respectively.
    \item Section \ref{sec: Implementation details} provide results detailing our choice in the use of {\tt stan} as our sampling algorithm over a Geodesic Monte Carlo implementation.
    \item Section \ref{sec: Simulated Data Examples} provides simulated data examples for assessing the quality of the model's ability to adequately recognize degree of separability for a data generating processes' covariance in static and seasonally dynamic covariances in Sections \ref{subsec: static covariance simulated data example} and \ref{subsec: seasonally dynamic covariance simulated examples}, respectively.
    \item Section \ref{sec: real data examples} provides real data applications of our model in a static case to the Wisconsin Breast Cancer dataset in Section \ref{subsec: Wiscosnin Breast Cancer} and a dynamic implementation for the analysis of coefficients in a Bayesian regression of continental US climate data corresponding to annual cyclic patterns in Section \ref{sec: Climate Analysis}.
    \item Section \ref{sec: conclusion} provides potential future directions of work, and Section \ref{sec: Computational results} provide supplementary details for efficient Hamiltonian Monte Carlo implementation of the static and seasonally dynamic parameterizations in Sections \ref{subsec: Static covariance computational results} and \ref{subsec: computational details seasonally dynamic covariance}, respectively.
\end{itemize}

\section{Mathematical Background} \label{sec: Mathematical Background}
\subsection{Bayesian Inference} \label{subsec: Bayesian Inference}
Bayesian inference aims to characterize the latent structure of a parameter space $\Gamma$ given a collection of observations $Y = \{y_{1},\ldots, y_{n}\}$. We do so assuming a prior distribution in our parameter space $p(\Gamma)$ and the likelihood of our data parameterized by $L(y; \Gamma)$. The goal of Bayesian inference is to characterize the posterior distribution of $\Gamma$ given $Y$ according to Bayes theorem
\[
P(\Gamma \vert Y) = \frac{L(Y;\Gamma) P(\Gamma)}{\int_{\gamma} L(Y;\gamma) P(\gamma) d\gamma },
\]
however, the calculation of the integral in the denominator is often intractable. As such, posterior inference often resolves to using computational techniques such as MCMC sampling or making approximations with known distributions such as variational inference.

MCMC sampling is asymptotically exact and is generally referred to as the gold standard for computational inference. Historically, MCMC inference is characterized by Gibbs \cite{gelfand2000gibbs} and random walk samplers. Gibbs sampling makes use of sampling from the full conditional posteriors. That is, by sampling from $P(\beta \vert Y, \Gamma_{-\beta})$. This, however, requires either prior distributions that are conditionally conjugate with the likelihood or analytically tractable expansions that are conditionally conjugate. Often these impose restrictions which limit expressivity of the prior and are difficult to derive. Random walk samplers, which are not constrained by the necessity of conditional conjugacy, are often slow to draw independent samples from the posterior distribution.

Hamiltonian Monte Carlo \cite{neal2011mcmc} is an advanced MCMC technique which extends random walk samplers by leveraging gradient information combined with updates defined by energy-preserving symplectic dynamics \cite{hofer2012symplectic} to build an ergodic Markov chain, which combined with a Metropolis correction gives asymptotically exact inference.

Mathematically, given a posterior distribution up to normalization, $P^{*}(q_{t} \vert Y) = L(Y ; q_{t}) P(q_{t})$, Hamiltonian Monte Carlo works by introducing an auxiliary latent variable $p_{t}$ with distribution $K(p_{t})$, and expressing a tradeoff between $p_{t}$ and $q_{t}$ through the time dependent coupled energy preserving differential equations
\begin{align}
    H(q_{t},p_{t}) &= P^{*}(q_{t} \vert Y) + K(p_{t}) \label{eq: Hamiltonian} \\
    \frac{\partial q_{t}}{\partial t} &= \frac{\partial H}{\partial p_{t}} \label{eq: Position Updates} \\
    \frac{\partial p_{t}}{\partial t} &= - \frac{\partial H}{\partial q_{t}}. \label{eq: Velocity Updates}
\end{align}

The differential equations (\ref{eq: Position Updates}-\ref{eq: Velocity Updates}) describe Hamilton's equations of motion. The key point is that under perfect evolution through $t$, Hamilton's equations give symplectic updates of $H$. That is, $H(q_{t},p_{t}) = H(q_{0}, p_{0})$. However, practical implementation often does not allow analytic solutions to Hamilton's equations, wherein the resolution is the leapfrog integrator (\cite{hut1995building}), where dynamics are discretized as:
\begin{align*}
    q_{t + \frac{1}{2}} &\rightarrow q_{t} + \frac{\epsilon}{2} \nabla_{p_{t}} H(q_{t}, p_{t}) \\
    p_{t + 1} &\rightarrow p_{t} +  \epsilon \nabla_{q_{t + \frac{1}{2}}} H(q_{t + \frac{1}{2}}, p_{t}) \\
    q_{t + 1} &\rightarrow q_{t + \frac{1}{2}} + \frac{\epsilon}{2} \nabla_{p_{t + 1}} H(q_{t + \frac{1}{2}}, p_{t + 1}) 
\end{align*}
for $t \in \{0,\ldots, T\}$. Trajectories under the leapfrog integrator are then accepted according to the Metropolis ratio:
\[
\alpha(q_{0}, q_{T}) = \min(1, \frac{\exp \big(-H(q_{T}, p_{T}) \big)}{\exp \big(-H(q_{0}, p_{0}) \big)}
\]
as $\epsilon \rightarrow 0$, the trajectories converge to symplectic trajectories, yielding a high $\alpha$ potentially at the cost of high autocorrelation for a choice of $T$ too small. Large choices of $\epsilon$ yield trajectories which yield a potentially too small $\alpha$ at the benefit of being highly uncorrelated. The choice of these tuning parameters then often comes down to optimizing based on a apriori choice of $\alpha$, which for sufficiently uncorrelated samples is typically $\approx .8$ in off the shelf samplers.
\subsection{Tensor Normal Distributions} \label{subsection: Tensor Normal}
The tensor normal model is a model used in the analysis of multi-indexed Gaussian data. That is, Gaussian data which may itself be reshaped into the form $\mathcal{X} \in \mathbb{R}^{\times_{i = 1}^{D} d_{i}}$. In particular, construction of the tensor normal class can be understood in terms of the Tucker product \cite{hitchcock1927expression,hoff2011separable}: for a tensor $\mathcal{A} \in \mathbb{R}^{\times_{i = 1}^{D}}$, and matrix $B \in \mathbb{R}^{d_{i} \times d_{i}}$ the Tucker product is defined as:
    \begin{equation} \label{eq: Tucker Product}
        (\mathcal{A} \times_{i} B)[k_{1}, \ldots, q_{i}, \ldots,  k_{D}] := \sum_{k_{i} = 1}^{d_{i}} \mathcal{A}[k_{1}, \ldots, k_{i}, \ldots, k_{D}] B[k_{i}, q_{i}].
    \end{equation}
    Alternatively for a collection of matrices $\mathbf{B} = \{B_{1}, \ldots, B_{D}\}$, the Tucker product between $\mathcal{A}$ and $\mathbf{B}$ can be stated directly as:
    \begin{equation} \label{eq: Full Tucker Product}
        (\mathcal{A} \circ \mathbf{B})[q_{1}, \ldots, q_{D}] := \sum_{k_{1} = 1}^{d_{1}} \cdots \sum_{k_{D} = 1}^{d_{D}} \mathcal{A}[k_{1}, \ldots, k_{D}]B_{1}[k_{1}, q_{1}] \cdots B[k_{D}, q_{D}].
    \end{equation}
    The interpretation of this tensor product is that the constituent matrices $\mathbf{B}$ act 'mode-wise' on $\mathcal{A}$. For example, in the 2-way case:
    \begin{equation} \label{eq: two way Tucker}
        (\mathcal{A} \circ \mathbf{B})[q_{1},q_{2}] := \sum_{k_{1} = 1}^{d_{1}} \sum_{k_{2} = 1}^{d_{2}} \mathcal{A}[k_{1}, k_{2}] B_{1}[k_{1}, q_{1}] B_{2}[k_{2}, q_{2}] = B_{1} \mathcal{A} B_{2}.
    \end{equation}
    A key point of utility in the use of the Tucker product is the ability to compute $(\mathcal{A} \circ \mathbf{B})$ sequentially through:
    \begin{equation} \label{eq: Sequential Tucker Operations}
    (\mathcal{A} \circ B) = \big(\big((\mathcal{A} \times_{1} B_{1}) \times_{2} B_{2} \big) \times_{3} B_{3}) \cdots \times_{i} B_{i}) \times \cdots \times_{D} B_{D}\big).
    \end{equation}
    To understand the convenience of this operation in terms of the normal distribution, the tensor normal itself is constructed through the assumption that element-wise, $\mathcal{A}$ is standard normal, and $\mathbf{B}_{i} \in \mathcal{P}^{+}(d_{i})$, wherein 
    \begin{equation} \label{eq: Tensor Normal Construction}
        \mathcal{T} \sim TN(\mathcal{M}, \{\Sigma_{1}, \ldots, \Sigma_{D}\}) \implies T \stackrel{D}{=} \mathcal{A} \circ \{ \Sigma_{1}^{\frac{1}{2}}, \ldots, \Sigma_{D}^{\frac{1}{2}}\} + \mathcal{M}. 
    \end{equation}
    We can produce an identical construction through the Cholesky factorization by instead letting $\{L_{1}, \ldots, L_{D}\} = \{\mathcal{L}(\Sigma_{1}), \ldots, \mathcal{L}(\Sigma_{D})\}$, where instead:
    \begin{equation} \label{eq: Tensor Normal Construction -- Cholesky}
        \mathcal{T} \sim TN(\mathcal{M}, \{\Sigma_{1}, \ldots, \Sigma_{D}\}) \implies T \stackrel{D}{=} \mathcal{A} \circ \{ L_{1}, \ldots, L_{D}\} + \mathcal{M}. 
    \end{equation}
    Where by \cite{kolda2006multilinear}, we can directly observe the tensor normal distribution is then a special case of the vector normal:
    \begin{equation} \label{eq: Tensor Normal vectorization -- Cholesky}
        vec(T) \stackrel{D}{=} \otimes_{i = D}^{1} L_{i} vec(\mathcal{A}) + vec(\mathcal{M}) \sim N(vec(M), \otimes_{i = D}^{1} L_{i}L_{i}^{T})
\end{equation}
where for $A \in \mathbb{R}^{d_{1} \times d_{1}}$, $B \in \mathbb{R}^{d_{2} \times d_{2}}$, $A \otimes B \in \mathbb{R}^{d_{1} d_{2} \times d_{1} d_{2}}$ denotes the Kronecker product:
\begin{equation} \label{eq: Kronecker Product}
    A \otimes B = \begin{pmatrix}
        a_{11} B &\cdots &a_{1,d_{1}}B \\
        \vdots & \ddots & \vdots \\
        a_{d_{1},1}B & \cdots & a_{d_{1},d_{1}} B
    \end{pmatrix}
\end{equation}
however, it's clear that in it's most general form, the vector normal distribution is computationally limited through large matrix multiplications for an arbitrary covariance. Whereas for data which admits a multi-index array structure, greater scalability can be achieved by leveraging the smaller sequential operations from (\ref{eq: Sequential Tucker Operations}) without a significant loss of flexibility as say an assumption of diagonality.

However, it is clear from the Kronecker structured covariance that the tensor normal distribution sacrifices some flexibility for scalability. Recall that the Cholesky decomposition corresponds to a specific choice of conjugate axes for the ellipsoid defined by the set $\{y: y^{t} (A\otimes B) y = 1\}$. Informally, the shape of this ellipsoid can be characterized by the component-wise eccentricity of the ellipsoid. In other words, the principal axes correspond to the eigenvectors of $A \otimes B$. SVD is known to be distributed on Kronecker products \cite{tucker1966some}:
\[
A \otimes B = (U_{A} \otimes U_{B})(D_{A} \otimes D_{B})(U_{A} \otimes U_{B}).
\]
Let $Q = C \oslash D$ denote the Hadamard division between $C,D \in \mathbb{R}^{n \times p}$, which is defined element-wise by:
\[
Q_{i,j} = \frac{C_{i,j}}{D_{i,j}}.
\]
Then note that
\[
(U_{A}\otimes U_{B}) [\cdot,d_{2}(j - 1)+  i] \oslash \big(U_{A}[\cdot,i] \otimes 1_{d_{2}}) = c_{i} 1_{d_{1}\cdot d_{2}}
\]
where $c_{i} \in \mathbb{R}$ for all $i,j \in \{1,\ldots,d_{1}\}\times \{1,\ldots, d_{2}\}$. Depending only on the index $i$, there are then only $d_{1}$ distinct principal directions that generate the ellipsoid $y^{t} (A \otimes B) y = 1$ up to scaling out of the total dimensions $d_{1} \cdot d_{2}$. These repetitions $d_{2}$ of each principal direction manifest in the repetitions of duplication and scaling found within the conjugate axes defined by $B$, scaled by the elements of $A$. These duplications are only distinct in their scaling, as such the shape as defined by the eccentricity is not distinct in these repetitions. 

In Section \ref{subsec: Enhancing the Tensor Normal}, we identify the exact sources of structure present within a Cholesky factorization of a tensor normal covariance and demonstrate that one such source can be completely eliminated without loss of analytic tractability of the likelihood through a simple modification introduced in the next section.

\section{Extending the Multiway Parameterization} \label{sec: extending the multi-way parameterization}
\subsection{Differential Geometry of Cholesky Factors} \label{sec: Differential Geometry of Cholesky Factors}
Let $\mathcal{P}^{+}(d)$ denote SPD matrices of $d \times d$ full rank, $\mathcal{L}^{+}(d)$ denote Cholesky factors of $d \times d$ SPD matrices, and $\lfloor \mathcal{L}^{+}\rfloor(d)$, $\mathbb{D}(\mathcal{L}^{+})(d)$ denote correspondingly the spaces of strictly lower triangular and diagonal components of Cholesky factors, respectively. We use $\mathcal{L}(\Sigma)$ to refer to the lower triangular Cholesky factor of $\Sigma$, note that $\mathcal{P}^{+}(d)$ is in bijective correspondence with $\mathcal{L}^{+}(d)$ and $\Sigma = \mathcal{L}(\Sigma) \big[ \mathcal{L}(\Sigma) \big]^{T}$. Let $\lfloor \mathcal{L}(\Sigma) \rfloor$ and $\mathbb{D}(\mathcal{L}(\Sigma))$ be strictly lower triangular and diagonal entries of $\mathcal{L}(\Sigma)$. If $\mathbf{I} = \{i_{1}, \ldots, i_{k}\}$ such that $i_{j} \in \mathbb{N}$, let $\mathcal{P}^{+}(\mathbf{I})$ denote the set of Kronecker structured covariance matrices as
\[
\mathcal{P}^{+}(\mathbf{I}) := \{\otimes_{j \in \mathbf{I}} A_{j} \vert A_{j} \in \mathcal{P}^{+}(i_{j})\}
\]
and correspondingly the set of Kronecker structured lower triangular Cholesky factors as
\[
\mathcal{L}^{+}(\mathbf{I}) := \{ \otimes_{j \in \mathbf{I}} \mathcal{L}(A_{j}) \vert \mathcal{L}(A_{j}) \in \mathcal{L}^{+}(i_{j})\}.
\]

A topological manifold is a second countable Hausdorff space which is locally holomorphic to the Euclidean space. For a topological manifold $\mathcal{M}$, the tangent space at $q \in \mathcal{M}$, $T_{q} \mathcal{M}$ is defined by the equivalence class of curves parameterized by the space
\[
T_{q}\mathcal{M} := \{ \gamma'(0): \gamma(0) = q\}.
\]

By endowing the tangent space of a topological manifold with a continuously varying bi-invariant metric, $g$, the paired structure $(\mathcal{M}, g)$ then becomes a Riemannian manifold. The key benefit of Riemannian manifolds with regard to this article is that they allow the computation of angles between tangent vectors and thereby allow the construction of distances between points, and importantly the construction of geometrically informed averages on nonlinear spaces, known as Fr\'echet means.

 The Riemannian geometry of the Cholesky factors of the SPD matrices was recently investigated in \cite{lin2019riemannian}. In particular, they derive the topological manifold structure of $\mathcal{L}^{+}(d)$, and endow the space with the following Riemannian metric:
\[
g_{L}(U,V) = tr(\mathbb{D}(U) \mathbb{D}(L)^{-1} \mathbb{D}(V) \mathbb{D}(L)^{-1}) + tr(\lfloor U\rfloor \lfloor V \rfloor),\quad U,V \in \mathcal{T}_{L} \mathcal{L}^{+}(d).
\]
Note that this metric should be viewed as endowed with the diagonal and strictly lower triangular components with a product manifold metric, where the diagonal is equipped with the affine-invariant metric and the strictly lower triangular entries are equipped with the Euclidean metric.
For $X \in \mathcal{T}_{L} \mathcal{L}^{+}(d)$, the corresponding geodesic at $L$ in the direction of $X$ under this metric are given by:
\begin{equation}\label{eq: Cholesky geodesic}
    \gamma_{L,X}(t) = \lfloor L \rfloor + t \lfloor X \rfloor + \exp\big(t \mathbb{D}(x) \mathbb{D}(L)^{-1} \big).
\end{equation}

For $L_{1}, L_{2} \in \mathcal{L}^{+}(d)$, the corresponding geodesic distance between $L_{1}$ and $L_{2}$ is given by:
\begin{equation} \label{eq: Cholesky distance}
    d_{\mathcal{L}^{+}(d)}(L_{1}, L_{2}) = \big[\| L_{1} - L_{2}\|_{F}^{2} + \| \log \mathbb{D}(L_{1}) - \log \mathbb{D}(L_{2})\|^{2} \big]^{\frac{1}{2}}.
\end{equation}

For a finite sample $\{A_{1}, \ldots, A_{n}\}$ belonging to a Riemannian manifold $(\mathcal{M}, g)$ with corresponding geodesic distance $d(A,B)$ for $A,B \in \mathcal{M}$, the Fr\'echet mean over $\mathcal{M}$ is defined as the sum of squares minimization:
\begin{equation} \label{eq: Frechet mean}
    \mu_{g} = \arg \min_{x \in \mathcal{M}} \sum_{i = 1}^{n} d^{2}(x, A_{i}).
\end{equation}

Under the log-Cholesky metric, for $\{L_{1}, \ldots, L_{n}\} \in \mathcal{L}^{+}(d)$, the corresponding Fr\'echet mean over $\mathcal{L}^{+}(d)$ is then given by the quantity
\begin{equation} \label{eq: cholesky mean}
\mathbb{E}_{n}(L_{1}, \ldots, L_{n}) = \frac{1}{n} \sum_{i = 1}^{n} \lfloor L_{i} \rfloor + \exp \big\{ \frac{1}{n} \sum_{i = 1}^{n} \log \mathbb{D}(L_{i}) \big\}.
\end{equation}
Our first observation is to note that the Fr\'echet mean of a finite sample of multiway Cholesky factors is, in general, no longer multiway.

\begin{theorem} \label{thm: cholesky mean non kronecker}
    Let $L^{\dagger}$ be the Cholesky factor such that:
    \begin{equation} \label{eq: L dagger}
    L^{\dagger} =  \sum_{i = 1}^{n} \lfloor \mathcal{L}(A_{i}) \rfloor + \exp \big\{ \sum_{i = 1}^{n} \log \mathbb{D}(\mathcal{L}(A_{i})) \big \}
    \end{equation}
    for $A_{i} \in \mathcal{P}^{+}(\mathbf{I})$, then $L^{\dagger} \not \in \mathcal{L}^{+}(\mathbf{I})$, while $\mathbb{D}(L^{\dagger}) \in \mathbb{D}(\mathcal{L}^{+})(\mathbf{I})$.

    \begin{proof}
        For the sake of proof suppose that $L^{\dagger} \in \mathcal{L}^{+}(\mathbf{I})$. Then $\exists A,B$ with $A \in \mathbb{D}^{+}(\mathbf{I})$ and $B \in \lfloor \mathcal{L}^{+} \rfloor (\mathbf{I})$ such that $L^{\dagger} = A + B$.  Now note, if $A_{i} \in \mathcal{P}^{+}(\mathbf{I})$ for $i \in \{1,\ldots, k\}$, then 
        \begin{equation}
            \log \mathbb{D}(\mathcal{L}(A_{t})) = \sum_{ t =1 }^{k} \otimes_{j = 1}^{k}(1_{j = t} log(\mathbb{D}({A_{t}^{(j)}})) + 1_{j \neq t} I_{j}).
        \end{equation}
        So we can then write
        \begin{align}
            \sum_{i = 1}^{n} \log \mathbb{D}(\mathcal{L}^{+}(A_{i})) &= \sum_{i = 1}^{n} \sum_{ t =1 }^{k} \otimes_{j = 1}^{k}(1_{j = t} log(\mathbb{D}({A_{i_{t}}^{(j)}})) + 1_{j \neq t} I_{j}) \\
            &=  \sum_{ t =1 }^{k} \otimes_{j = 1}^{k}(1_{j = t} \sum_{i = 1}^{n}log(\mathbb{D}({A_{i_{t}}^{(j)}})) + 1_{j \neq t} I_{j}).
        \end{align}
        The summands are multiplicatively commutative, and so by proposition 2.3 of (\cite{hall2013lie}) and using the fact that for a continuous matrix valued function, $f(A \otimes I) = f(A) \otimes I$:
        \[
        \exp(\sum_{ t =1 }^{k} \otimes_{j = 1}^{k}(1_{j = t} \sum_{i = 1}^{n}log(\mathbb{D}({A_{i_{t}}^{(j)}})) + 1_{j \neq t} I_{j})) = \otimes_{j = 1}^{k} \big(\exp[1_{j = t} \sum_{i = 1}^{n}log(\mathbb{D}({A_{i_{t}}^{(j)}}))] + 1_{j \neq t} I_{j} \big),
        \]
        hence, $\exp \{ \sum_{i = 1}^{n} \log \mathbb{D}(\mathcal{L}^{+}(A_{i}))\} \in \mathbb{D}^{+}(\mathbf{I}) = \mathbb{D}(\mathcal{L}^{+})(\mathbf{I})$. 
        
        To show $\frac{1}{n} \sum_{i = 1}^{n} \lfloor \mathcal{L}^{+}(A_{i}) \rfloor \not \in \lfloor \mathcal{L}^{+} \rfloor (\mathbf{I})$, without loss of generality suppose $\#(\mathbf{I}) = 2$. Further suppose that $\mathcal{L}^{+}(A_{i}) = \mathcal{L}^{+}(A_{i}^{(1)})\otimes \mathcal{L}^{+}(A_{i}^{(2)})$. Then observe
        \begin{equation} \label{eq: full kronecker cholesky}
        \mathcal{L}^{+}(A_{i}) = \big(\lfloor \mathcal{L}^{+}(A_{i}^{(1)}) \rfloor + \mathbb{D}(\mathcal{L}^{+}(A_{i}^{(1)}) \big) \otimes \big(\lfloor \mathcal{L}^{+}(A_{i}^{(2)}) \rfloor + \mathbb{D}(\mathcal{L}^{+}(A_{i}^{(2)}) \big).
        \end{equation}
        It must then be necessarily true if $B \in \lfloor \mathcal{L}^{+} \rfloor (\mathbf{I})$ that $\exists C \in \mathcal{L}^{+}(d_{1}), D \in \mathcal{L}^{+}(d_{2})$ such that:
        \begin{equation} \label{eq: full kronecker cholesky pt 2}
        B = \lfloor C \rfloor \otimes \lfloor D \rfloor + \lfloor C \rfloor \otimes \mathbb{D}(D) + \mathbb{D}(C) \otimes \lfloor D \rfloor.
        \end{equation}
        Note that each of these components contribute independently to the strictly lower triangular elements of the lower triangular blocks, the strictly lower triangular elements of the lower triangular blocks, and the diagonal elements of the strictly lower triangular blocks, respectively.

        Let $\Psi$ denote any of these corresponding blocks of $P + Q$ for $P,Q \in \mathcal{L}^{+}(\mathbf{I})$ such that $P \neq Q$. Letting $I$ denote the non-zero elements of $P$ and $Q$, if $A + B$ were Kronecker structured, it must necessarily follow that for $\{i_{1},j_{1}\}, \{s_{1},t_{1}\} \in I$ such that 
        \[
        i_{1}, s_{1}, j_{1}, t_{1} \in \{md_{2} + 1, \ldots, d_{2}(m + 1)\}
        \]
        where $m \in \{0, \ldots, d_{1} - 1\}$:
        \[
        \frac{\Psi_{i_{1},j_{1}}}{\Psi_{s_{1},t_{1}}} = \gamma =  \frac{P_{i_{1},j_{1}} + Q_{i_{1},j_{1}}}{P_{s_{1},t_{1}} + Q_{s_{1},t_{1}}}
        \]
        letting $i_{2}, j_{2}, s_{2}, t_{2} = i_{1}, j_{1}, s_{1}, t_{1} + d_{2}$, assuming the corresponding element exists and is non-zero, it necessarily leads to
        \[
        \frac{\Psi_{i_{2},j_{2}}}{\Psi_{s_{2},t_{2}}} = \gamma =  \frac{P_{i_{2},j_{2}} + Q_{i_{2},j_{2}}}{P_{s_{2},t_{2}} + Q_{s_{2},t_{2}}}.
        \]
        Assuming $P, Q$ are the corresponding blocks of the matrices $A \otimes B$, $C \otimes D$, respectively,
        \begin{align*}
            P_{i_{1}, j_{1}} &= a_{1} b_{i,j}, \quad P_{i_{2}, j_{2}} = a_{2} b_{i,j}, \quad Q_{i_{1}, j_{1}} = c_{1} d_{i,j}, \quad Q_{i_{2}, j_{2}} = c_{2} d_{i,j} \\
            P_{s_{1}, t_{1}} &= a_{1} b_{s,t}, \quad P_{s_{2}, t_{2}} = a_{2} b_{s,t}, \quad Q_{s_{1}, t_{1}} = c_{1} d_{s,t}, \quad Q_{s_{2}, t_{2}} = c_{2} d_{s,t}
        \end{align*}
        it would then follow:
        \[
        \frac{a_{1} b_{i,j} + c_{1} d_{i,j}}{a_{1} b_{s,t} + c_{1} d_{s,t}} = \frac{a_{2} b_{i,j} + c_{2} d_{i,j}}{a_{2} b_{s,t} + c_{2} d_{s,t}}.
        \]
        However, this does not in general hold for arbitrary SPD matrices $A,B,C,D$. The general multiway case holds analogously.
    \end{proof}
\end{theorem}

Of importance is to note that it does not generally hold that Fr\'echet means on separable manifolds are always non-separable. For example, if $T_{i} \in \mathcal{P}^{+}(\mathbf{I})$, $\log(T_{i}) = \sum_{j = 1}^{k} \otimes_{t = 1}^{k} (1_{t = j} \log(A_t) + 1_{t \neq j} I_{t})$. Given a finite sample $T_{1}, \ldots, T_{n} \in \mathcal{P}^{+}(d)$, the log-Euclidean Fr\'echet mean is given by \cite{utpala2022differentially} as the following expression
\[
\mu_{LE}(\{T_{i}\}_{i = 1}^{n}) = \exp(\frac{1}{n} \sum_{i = 1}^{n} \log (T_{i}))
\]
and as the summands of the exponent are multiplicatively commutative, it immediately holds $\mu_{LE}(\{T_{i}\}_{i = 1}^{n}) \in \mathcal{P}^{+}(\mathbf{I})$. Moreover, while the Affine-Invariant metric has no closed-form Fr\'echet mean, we found a similar result numerically for it.

As an analog to Corollary 12 of \cite{lin2019riemannian}, even though the result of Theorem \ref{thm: cholesky mean non kronecker} states that the Fr\'echet mean in multiway Cholesky space does not yield something which is Kronecker structured, the log determinant is still the geometric mean of the determinants,

\begin{proposition} \label{thm: expected cholesky determinant}
    Let $L^{\dagger}$ be defined as in Theorem \ref{thm: cholesky mean non kronecker}, then 
    \[
    \det L^{\dagger} = \exp( \sum_{i = 1}^{n} \sum_{j = 1}^{k} d_{-j} \log \det L_{i}^{(j)}).
    \]
    \begin{proof}
        The result is a direct extension of Corollary 13 in \cite{lin2019riemannian}, where we additionally recognize $\log \vert L_{i} \vert = \sum_{j = 1}^{k} d_{-j} \log \vert L_{i}^{(j)} \vert$.
    \end{proof}
\end{proposition}
 Here, in particular, the emphasis is that $\mathbb{E}_{n}(L_{1}, \ldots, L_{n}) \not \in \mathcal{L}^{+}(\mathbf{I})$. The key point is that the diagonal still exists in Kronecker space; hence the determinant of the Fr\'echet mean is expressible as a decomposition in terms of determinants of the components. However, because the strictly lower triangular part cannot be appropriately expressed in Kronecker product form, it cannot exist as a single Kronecker product.

\subsection{Enhancing The Tensor Normal} \label{subsec: Enhancing the Tensor Normal} 
The classical covariance parametrization of the $d$ dimensional multivariate normal density is given by:
\begin{equation} \label{eq: Multinormal covariance}
    f(x \vert \Sigma, \mu) = \frac{1}{(2\pi)^{d/2} \vert \Sigma \vert^{\frac{1}{2}}} \exp(-\frac{1}{2} (x - \mu) \Sigma^{-1} (x - \mu)).
\end{equation}
Equivalently, representing this this density with the precision matrix,  $\Theta = \Sigma^{-1}$, yields
\begin{equation} \label{eq: Multinormal precision}
    f(x \vert \Sigma, \mu) = \frac{\vert \Theta \vert^{\frac{1}{2}}}{(2\pi)^{d/2} } \exp(-\frac{1}{2} (x - \mu) \Theta (x - \mu)).
\end{equation}
Alternatively applying the Cholesky decomposition of the precision matrix, $\Theta = L_{\theta}L_{\theta}^{T}$, the density becomes
\begin{equation} \label{eq: Multinormal cholesky precision}
    f(x \vert L_{\theta}, \mu) = \frac{\vert L_{\theta} \vert}{(2\pi)^{d/2}} \exp\big(-\frac{1}{2} (x - \mu) L_{\theta} L_{\theta}^{T} (x - \mu) \big).
\end{equation}
Now let $T = \{L_{1}, \ldots, L_{n}\}$ such that each $L_{i} \in \mathcal{L}^{+}(\mathbf{I})$ such that $\#(\mathbf{I}) = 2$. Using the results of the previous section, let $L^{\dagger}$ have the same meaning as in (\ref{eq: L dagger}):
\begin{equation} \label{eq: Multinormal cholesky precision mean}
    f(x \vert T, \mu) = \frac{ \vert L^{\dagger}  \vert}{(2\pi)^{d/2}} \exp\big(-\frac{1}{2} (x - \mu) L^{\dagger} (L^{\dagger})^{T} (x - \mu) \big).
\end{equation}
Which by Proposition \ref{thm: expected cholesky determinant} yields:
\begin{align}
    f(x \vert T, \mu) &= \frac{\exp( \sum_{i = 1}^{n} \sum_{j = 1}^{k} d_{-j} \log \det L_{i}^{(j)}).}{(2\pi)^{d/2}} \exp\big(-\frac{1}{2} (x - \mu)(x - \mu)^{T} \\
    &\big[ \sum_{i = 1}^{n} \lfloor \mathcal{L}(A_{i}) \rfloor + \exp \big\{ \sum_{i = 1}^{n} \log \mathbb{D}(\mathcal{L}(A_{i})) \big \} \big] 
    \big[\sum_{i = 1}^{n} \lfloor \mathcal{L}(A_{i}) \rfloor + \exp \big\{ \sum_{i = 1}^{n} \log \mathbb{D}(\mathcal{L}(A_{i})) \big \}\big]^{T} \big).
\end{align}

Note that we can further simplify this likelihood to involve only the constituent components of the Kronecker product. Specifically, in the case $\#(\mathbf{I}) = 2$, we can say $A_{i} = A_{1}^{(i)} \otimes A_{2}^{(i)}$, in which case $\mathcal{L}(A_{i}) = \mathcal{L}(A_{1}^{(i)} \otimes A_{2}^{(i)}) = \mathcal{L}(A_{i}^{(1)}) \otimes \mathcal{L}(A_{i}^{(2)})$. By the same reasoning as (\ref{eq: full kronecker cholesky}) - (\ref{eq: full kronecker cholesky pt 2}):
\begin{align*}
    \mathcal{L}(A_{1}^{(i)}) \otimes \mathcal{L}(A_{2}^{(i)}) &= \big[ \lfloor \mathcal{L}(A_{1}^{(i)}) \rfloor + \mathbb{D}(\mathcal{L}(A_{1}^{(i)}))\big] \otimes \big[\lfloor \mathcal{L}(A_{2}^{(i)}) \rfloor + \mathbb{D}(\mathcal{L}(A_{2}^{(i)}) ) \big]\\
    \implies \lfloor \mathcal{L}(A_{i}) \rfloor &= \lfloor \mathcal{L}(A_{1}^{(i)})\rfloor \otimes \lfloor \mathcal{L}(A_{2}^{(i)} \rfloor + \mathbb{D}(\mathcal{L}(A_{1}^{(i)}))\otimes \lfloor \mathcal{L}(A_{2}^{(i)})\rfloor + \lfloor \mathcal{L}(A_{1}^{(i)}) \rfloor \otimes \mathbb{D}(\mathcal{L}(A_{2}^{(i)}) )\\
    \mathbb{D}(\mathcal{L}(A_{i})) &= \mathbb{D}(\mathcal{L}(A_{1}^{(i)})) \otimes \mathbb{D}(\mathcal{L}(A_{2}^{(i)})).
\end{align*}
Moreover, it's straightforward to see that
\[
\big[\sum_{i = 1}^{n} \lfloor \mathcal{L}(A_{i}) \rfloor + \exp \big\{ \sum_{i = 1}^{n} \log \mathbb{D}(\mathcal{L}(A_{i})) \big \}\big]^{T} = \sum_{i = 1}^{n} \lfloor \mathcal{L}(A_{i}) \rfloor^{T} + \exp \big\{ \sum_{i = 1}^{n} \log \mathbb{D}(\mathcal{L}(A_{i})) \big \}\big].
\]

For brevity of this section, we will refrain from the full computation of the likelihood until the supplemental computational details are provided in Section \ref{sec: Computational results}.

This parameterization allows for a more general representation of Cholesky factors than the more straightforward argument of separability. In particular, as stated in Theorem \ref{thm: cholesky mean non kronecker}, a general Cholesky factor, $L$, can only be adequately represented with a multiway Cholesky factorization when it is strictly lower triangular elements then
\[
\frac{L_{i_{1}, j_{1}}}{L_{s_{1}, t_{1}}} = \frac{L_{i_{2}, j_{2}}}{L_{s_{2}, t_{2}}}.
\]

However, under our parameterization, we are able to circumvent this structural assumption, as observed in the following Lemma for the matrix normal case.
\begin{lemma} \label{lemma: strictly lower rep}
    Let $\Psi \in \mathbb{R}^{d_{1}d_{2} \times d_{1} d_{2}}$ be an arbitrary strictly lower triangular matrix such that 
    \[
    \Psi_{d_{2}r + v, d_{2}s + w} =0
    \]
    if $r > s$ or $w > v$. Then
    \[
    \arg\min_{A,B} \| \Psi - A\otimes B\|_{F} \in \lfloor \mathcal{L}^{+} \rfloor(d_{1}) \times \lfloor \mathcal{L}^{+}\rfloor (d_{2}).
    \]
    \begin{proof}  Assuming $A,B \in \mathbb{R}^{d_{1}\times d_{1}} \times \mathbb{R}^{d_{2} \times d_{2}}$ are unconstrained, this is a direct result of the following calculations:
    \begin{align*}
        &\| \Psi - A \otimes B\|_{F}^{2} = \sum_{i = 1}^{d_{1}d_{2}}\sum_{j = 1}^{d_{1}d_{2}} \vert \Psi_{i,j} - \big[A \otimes B \big]_{i,j} \vert^{2}\\
        &= \sum_{r = 1}^{d_{1} - 1} \sum_{s = 1}^{d_{1} - 1} \sum_{v = 1}^{d_{2}} \sum_{w = 1}^{d_{2}} \vert \Psi[d_{2}r + v, d_{2}s + w] - \big[A \otimes B \big][d_{2}r + v, d_{2}s + w] \vert^{2} \\
        &= \sum_{r \geq s} \sum_{v \geq w} \vert \Psi[d_{2}r + v, qs + w] - \big[A \otimes B\big][d_{2}r + v, d_{2}s + w] \vert^{2} \\
        &+ \sum_{r \geq s} \sum_{w < v} \vert A_{r,s} B_{v,w}\vert^{2} + \sum_{r < s} \sum_{w \geq v} \vert A_{r,s} B_{v,w}\vert^{2} + \sum_{r \leq s} \sum_{w \leq v} \vert A_{r,s} B_{v,w}\vert^{2}.
    \end{align*}
    The last line follows from $(A \otimes B)_{d_{2}r + v, d_{2} s + w} = A_{r,s} B_{v,w}$ and merely penalizes non-strictly lower triangular elements of A and B. Hence, $\| \Psi - A \otimes B\|_{F}^{2}$ is minimized when $A$ and $B$ are lower triangular. 
    \end{proof}
\end{lemma}

By the works of \cite{pitsanis1997kronecker} and \cite{VanLoan1993}, it is analogous that then by iterating this process on the residuals of $\Psi - A\otimes B$ for strictly lower triangular $\Psi$, there exists a collection of matrices $\{A_{i}, B_{i}\}_{i = 1}^{r^{2}}$ where $r = \min\{d_{1}, d_{2}\}$ such that
\begin{equation} \label{eq: zero residual}
\| \Psi - \sum_{i = 1}^{r^{2}} A_{i} \otimes B_{i} \|_{F} = 0.
\end{equation}
We refer to this representation as a sum of Kronecker products as the Pitsianis-Van Loan (P-VL) decomposition.

Then analogously with respect to the general capability of representing an arbitrary Cholesky factorization of a precision matrix through a Fr\'echet mean of separable Cholesky factors, we observe the following lemma:
\begin{lemma}
    Let $L^{\dagger} = \mu_{LC}(\{\otimes_{i = 1}^{2} L_{i}^{(j)}\}_{j = 1}^{K})$ be the Cholesky Fr\'echet mean of a $K$-collection of multiway Cholesky factors where $K = \min\{d_{1}^{2},d_{2}^{2}\}$ such that $\mathbb{D}(\otimes_{i = 1}^{2} L_{i}^{(p)}) = \mathbb{D}(\otimes_{i = 1}^{2} L_{i}^{(q)})$ for $p \neq q$, then for arbitrary $L \in \mathcal{L}^{+}(d_{1}d_{2})$:
    \[
    \min_{L^{\dagger}}\|L^{\dagger} - L\|_{F} \leq \sqrt{ \sum_{r > s} \sum_{v > w}  L_{d_{2}r + v, d_{2}s + w}^{2}}  +  D_{2}(L).
    \]
    where $D_{2}(L)$ is the second largest diagonal of $L$.
    \begin{proof} By triangle inequality,
    \[
        \|L^{\dagger} - L\|_{F} \leq \|\lfloor L^{\dagger} \rfloor - \lfloor L \rfloor \|_{F} + \| \mathbb{D}(L^{\dagger}) - \mathbb{D}(L)\|_{F}
    \]
    First note that by iteratively minimizing the residuals of $\Psi - A\otimes B$ for $\Psi$ has the sparsity structure of Lemma (\ref{lemma: strictly lower rep}), through the argument of the P-VL decomposition, it immediately follows by a second application of triangle inequality:
    \begin{align*}
        \min_{L^{\dagger}}\| L^{\dagger} - L \|_{F} &\leq \sqrt{ \sum_{r > s} \sum_{v > w}  L_{d_{2}r + v, d_{2}s + w}^{2}}  +  \min_{L^\dagger} \sqrt{ \sum_{r \leq s} \sum_{v \leq w}  L_{d_{2}r + v, d_{2}s + w}^{2} - L^{\dagger}_{d_{2}r + v, d_{2}s + w}} \\
        &= \sqrt{ \sum_{r > s} \sum_{v > w}  L_{d_{2}r + v, d_{2}s + w}^{2}}
    \end{align*}
    where the second line follows for sufficiently large $K$ by equation (\ref{eq: zero residual}). By a secondary application of P-VL it also follows:
    \[
    \min_{L^{\dagger}}\|\mathbb{D}(L^{\dagger}) - \mathbb{D}(L)\| = \min_{D_{1}, D_{2}}\| \otimes_{i = 1}^{d} D_{i} - \mathbb{D}(L)\| \leq  D_{2}(L).
    \]
    \end{proof}
\end{lemma}

Observe the implication of the above Lemma is that we can arbitrarily represent strictly lower triangular matrices through an analogue of the P-VL decomposition given the matrix has an appropriate sparsity structure (specifically, the lower triangular sparsity structure introduced from multiway Cholesky factors).

\section{The Bayesian Model} \label{sec: The Bayesian Model}
The model of \cite{hoff2011separable} provides an excellent starting point for adequately structuring an appropriate Bayesian model for multiway data. Assuming that 
\[
y_{1}, \ldots, y_{n} \sim N(0, \otimes_{i = 1}^{D} \Sigma_{i}).
\]
Given a lack of scale identifiability in any of the components present in the data:
\[
(c A) \otimes (\frac{1}{c} B) = A \otimes B
\]
the authors propose to estimate the $\{\Sigma_{i}\}_{i = 1}^{D}$ by means of centering the prior expectations of the covariance:
\[
\Sigma_{i} \sim IW(d_{i} + 2, \frac{\gamma^{\frac{1}{D}}}{d_{i}} I_{d_{i}})
\]
such that $\gamma = tr(\frac{1}{n-1}\sum_{i = 1}^{n} y_{i} y_{i}^{T})$. The central point of this argument being that the prior centering of the covariance allows one to argue:
\begin{align*}
    \mathbb{E}[tr(\otimes_{i = 1}^{D} \Sigma_{i})] &= tr(\mathbb{E}[\otimes_{i = 1}^{D} \Sigma_{i}]) = 
    \prod_{i = 1}^{D} \frac{\gamma^{\frac{1}{D}}}{d_i}tr(I_{d_{i}}) = \gamma.
\end{align*}
Consequently, there is little information in the structure of the individual $\{\Sigma_{i}\}$ present in the data itself, but we can use the linearity of the trace to provide a motivating relationship between our prior expectations and the trace of the sample covariance matrix. Directly extending such a summation directly to the Fr\'echet mean would necessarily then require finding a parameterization for centering:
\[
\mathbb{E}[tr(\exp(D))] = \mathbb{E}[\exp(\vert D \vert)].
\]
In the next section, we instead parameterize $\exp(D_{1}) \otimes \exp(D_{2})$ directly with priors with only positive support.

\subsection{Centering Matrix Priors} \label{sec: Centering Matrix Priors}
In this section we will introduce several auxiliary results used in the construction of Bayesian models in Sections \ref{sec: Non-Dynamic} and \ref{sec: seasonally dynamic covariance}.
\begin{lemma}
    Suppose $c > 1$ and let $\alpha(c)$ be the solution to $f(x) = c\exp(-\frac{2}{x}) - 1 - \frac{1}{x}$. If $\alpha(c)$ is large enough to satisfy $\alpha(c) + \alpha^{2}(c) < \frac{\epsilon}{\exp(\frac{1}{\alpha}) - 1}$, then $\vert \alpha^{2} + \alpha - c\exp(2\psi_{0}(\alpha)) \vert < \epsilon$.
    \begin{proof}
        For $x > 0$, a known bound on the digamma function is given by:
        \begin{equation} \label{eq: digamma bounds}
            \psi_{0}(x) \in (\log(x) - \frac{1}{x}, \log(x) - \frac{1}{2x}).
        \end{equation}
        It follows:
        \[
        -c\exp(2\psi_{0}(x)) \in \big(-cx^{2} \exp(-\frac{1}{x}), -cx^{2} \exp(-\frac{2}{x})\big).
        \]
        Then we can bound 
        \[
        x^{2} + x - c\exp(2\psi_{0}(x)) \in \big(x^{2}(1 - c\exp(-\frac{1}{x})) + x, x^{2}(1- c\exp(-\frac{2}{x})) + x\big) = (l(x), u(x))
        \]
        For $x > 0$, $u(x) = 0$ occurs when
        \begin{equation} \label{eq: upper bound root}
        c\exp(-\frac{2}{x}) - 1 = \frac{1}{x}.
        \end{equation}
        Note that the right-hand side is positive and strictly decreasing in $x$, while the left-hand side is continuous and strictly increasing with corresponding limits $-1$ and $c-1$ as $x \rightarrow 0$ and $x \rightarrow\infty$, respectively, for any $c > 0$, so a solution must exist. Letting $\alpha$ be the solution to (\ref{eq: upper bound root}), it follows that 
        \begin{equation}
            w(\alpha) = u(\alpha) - l(\alpha) = \alpha^{2} c \big(\exp(-\frac{1}{\alpha}) - \exp(-\frac{2}{\alpha}) \big).
        \end{equation}
        Note that from the solution to (\ref{eq: upper bound root}), it must follow $c\exp(-\frac{2}{\alpha}) = \frac{1 + \alpha}{\alpha}$, and therefore 
        \[
        w(\alpha) = \alpha^{2}(\frac{1 + \alpha}{\alpha}\big(\exp(\frac{1}{\alpha}) - 1) \big) = (\alpha + \alpha^{2})(\exp(\frac{1}{\alpha}) - 1).
        \]
        From which if $\alpha(c)$ satisfies the conditions $w(\alpha) < \epsilon$ and $u(\alpha) = 0$, where we can deduce $\vert f(\alpha) \vert \leq u(\alpha) + w(\alpha) < \epsilon$.
    \end{proof}
\end{lemma}

\begin{proposition} \label{prop: independent diagonal parameterization}
    Let $D_{1} \in \mathbb{D}(d_{1})$, $D_{2} \in \mathbb{D}(d_{2})$ be independently parameterized by:
    \[
    D_{i}[j_{i}, j_{i}] \sim \Gamma(a_{i}, \exp( \psi_{0}(a_{i}) - \frac{\gamma_{D}}{2 d_{1}d_{2}})), \quad j_{i} \in \{1, \ldots, d_{i}\}
    \]
    such that $a_{i}^{2} + a_{i} - c_{i} \exp(2 \psi_{0}(a_{i}))  = \epsilon$ for $c_{i} = \frac{\sqrt{F_D}}{d_{i}} exp(-\gamma_D/ (2d_1d_2))$ for constant $F_{D}, \gamma_{D}$, then
    \begin{equation} \label{eq: expected log determinant gamma}
    \mathbb{E}[\log \vert D_{1} \otimes D_{2} \vert ] = \gamma_{D}
    \end{equation}
    and 
    \begin{equation} \label{eq: expected frobenius norm gamma}
    \mathbb{E}[\| D_{1} \otimes D_{2}\|_{F}^{2}] = (\sqrt{F_{D}} + \frac{d_{1}\epsilon}{ \exp(-\frac{\gamma_{D}}{2})\exp(2 \psi_{0}(a_{1}))})(\sqrt{F_{D}} + \frac{d_{2}\epsilon}{ \exp(-\frac{\gamma_{D}}{2}) \exp(2 \psi_{0}(a_{2}))}).
    \end{equation}
    \begin{proof}
        Note first that if $X \sim \Gamma(a,b)$, with rate $b$, it's known that
        \begin{equation} \label{eq: expected log gamma}
        \mathbb{E}[\log (X)] = \psi(a) - \log (b)
        \end{equation}
        and moreover
        \begin{equation} \label{eq: expected squared gamma}
            \mathbb{E}[X^{2}] = \frac{a + a^{2}}{b^{2}}.
        \end{equation}
        Now note that 
        \begin{equation} \label{eq: kronecker log determinant}
        \log \vert D_{1} \otimes D_{2} \vert = d_{2} \log \vert D_{1} \vert + d_{1} \log \vert D_{2} \vert
        \end{equation}
        and observe by  (\ref{eq: expected log gamma}):
        \[
        \log \vert D_{i}\vert = \sum_{j = 1}^{d_{i}} \log D_{i}[j,j] \implies \mathbb{E}[\log \vert D_{i} \vert ]= \sum_{j = 1}^{d_{i}} [\psi_{0}(a_{i}) - \psi_{0}(a_{i}) + \frac{\gamma_{D}^{\frac{1}{2}}}{d_{1}d_{2}}] = \frac{\gamma_{D}}{d_{-i}}.
        \]
        By linearity, it follows from equation (\ref{eq: kronecker log determinant}) that 
        \[
        \mathbb{E}[\log \vert D_{1} \otimes D_{2} \vert ] = \gamma_{D}. 
        \]
        To show that (\ref{eq: expected frobenius norm gamma}), note another convenient property of Kronecker products:
        \begin{equation} \label{eq: kronecker frobenius property}
        \|A \otimes B\|_{F}^{2} = \|A \|_{F}^{2} \| B \|_{F}^{2}
        \end{equation}
        for general matrices $A,B$. It follows that
        \[
        \| D_{1} \otimes D_{2} \|_{F}^{2} = \|D_{1} \|_{F}^{2} \|D_{2} \|_{F}^{2}. 
        \]
        Further observe that
        \[
        \mathbb{E}[\|D_{i}\|_{F}^{2}] = \frac{1}{\exp(- \frac{\gamma_{D}}{d_{1} d_{2}})}\sum_{j = 1}^{d_{i}} \frac{a_{i}^{2} + a_{i}}{\exp(2 \psi_{0}(a_{i}))}
        \]
        given $a_{i}^{2} + a_{i} - c_{i} \exp(2 \psi_{0}(a_{i})) < \epsilon$.  It then follows that
        \[
        \frac{1}{\exp(- \frac{\gamma_{D}}{d_{1} d_{2}})}\sum_{j = 1}^{d_{i}} \frac{a_{i}^{2} + a_{i}}{\exp(2 \psi_{0}(a_{i}))} = \frac{1}{\exp(- \frac{\gamma_{D}}{d_{1}d_{2}})} d_{i} (c_{i} + \frac{\epsilon}{\exp(2\psi_{0}(a_{i}))}).
        \]
        Given that
        \[
        \frac{1}{\exp(- \frac{\gamma_{D}}{d_{1}d_{2}})} d_{i} c_{i} = \sqrt{F_{D}}
        \]
        the result immediately follows by independence of $D_{1}$, $D_{2}$.
    \end{proof} 
\end{proposition}
Given that $\epsilon$ is arbitrary, we will resort to the notation of $\approx$ to signify a precision up to $\epsilon$. With this notation and Proposition \ref{prop: independent diagonal parameterization}, we can parameterize the strictly lower triangular entries as follows:
\begin{theorem} \label{thm: lower triangular expected frobenius norm}
    Let $L_{1}$, $L_{2}$ be positive definite Cholesky factors. Suppose $\mathcal{D}(L_{1}) \in \mathbb{D}(d_{1})$, $\mathcal{D}(L_{2}) \in \mathbb{D}(d_{2})$ be parameterized according to Proposition \ref{prop: independent diagonal parameterization}, further suppose the lower triangular entries are independently distributed according to
    \begin{align}
        \lfloor L_{1} \rfloor[i,j] &\sim N(0, \sigma^{2} = \beta) \\
        \lfloor L_{2}\rfloor[p,q] &\sim N(0, \sigma^{2} = \beta )
    \end{align}
    with
    \begin{equation} \label{eq: beta variance}
        \beta = \frac{-M_{1} \sqrt{F_D}(1+1/c) + \sqrt{F_{D}(1+1/c)^{2} + 4((M_{1})^{2}/c)*F_L))}}{(2*M_{1}^{2}/c)}
    \end{equation}
    $c = \frac{d_{1}(d_{1} - 1)}{d_{2}(d_{2} - 1)}$ and $M_{1} = \frac{d_{1}(d_{1} - 1)}{2}$, then 
    \[
    \mathbb{E}[\| \lfloor L_{1} \otimes L_{2} \rfloor  \|_{F}^{2}] \approx F_{L}.
    \]
    \begin{proof}
        First observe that for $L_{1} \in \mathcal{L}^{+}(d_{1})$, $L_{2} \in \mathcal{L}^{+}(d_{2})$,
        \[
        \lfloor L_{1} \otimes L_{2} \rfloor = \lfloor L_{1} \rfloor \otimes \mathcal{D}(L_{2}) + \mathcal{D}(L_{1}) \otimes \lfloor L_{2} \rfloor + \lfloor L_{1} \rfloor \otimes \lfloor L_{2} \rfloor = M_{1} + M_{2} + M_{3}.
        \]
        Observe $M_{1}, M_{2}, M_{3}$ respectively contribute structurally to the strictly lower triangular entries of the diagonal blocks, the diagonals of the strictly lower triangular blocks, and the strictly lower triangular components of the strictly lower triangular blocks. As such, letting $\odot$ denote the Hadamard product, then
        $ M_{i} \odot M_{j} = 0 \text{ for } i\neq j$ and note that in general, if $A \odot B = 0$ for arbitrary matrices $A,B$ compatible for such a multiplication, then
        $ \|A + B \|_{F}^{2} = \|A \|_{F}^{2} + \|B \|_{F}^{2}$.  It then follows that
        \[
        \| \lfloor L_{1} \otimes L_{2} \rfloor \|_{F}^{2} = \| \lfloor L_{1} \rfloor \otimes \mathcal{D}(L_{2}) \|_{F}^{2} + \| \mathcal{D}(L_{1}) \otimes \lfloor L_{2} \rfloor \|_{F}^{2} + \| \lfloor L_{1} \rfloor \otimes \lfloor L_{2} \rfloor \|_{F}^{2}
        \]
        and by (\ref{eq: kronecker frobenius property}),
        \[
        \| \lfloor L_{1} \otimes L_{2} \rfloor \|_{F}^{2} = \|\lfloor L_{1} \rfloor\|_{F}^{2} \|\mathcal{D}(L_{2})\|_{F}^{2} + \|\mathcal{D}(L_{1})\|_{F}^{2} \| \lfloor L_{2} \rfloor \|_{F}^{2} + \| \lfloor L_{1} \rfloor \|_{F}^{2} \| \lfloor L_{2} \rfloor\|_{F}^{2}.
        \]
        Further suppose that we have element-wise independent priors on $\lfloor L_{1} \rfloor$ as
        \[
        \lfloor L_{1} \rfloor[i,j] \sim N(0,\sigma^{2} = \xi^{2})
        \]
        then it follows that
        \[
        \mathbb{E}[\|\lfloor L_{1} \rfloor \|_{F}^{2}] = \frac{d_{1}(d_{1} - 1)}{2} \xi^{2}.
        \]
        Further assuming 
        \[
        \mathbb{E}[\| L_{1}\|_{F}^{2}] = c \mathbb{E}[\|L_{2}\|_{F}^{2}]
        \]
        then it is natural to assume 
        \[
        \lfloor L_{2} \rfloor[p,q] \sim N(0,\sigma^{2} = \frac{\xi^{2}}{c}),
        \]
        then it follows that
        \[
        \mathbb{E}[\|\lfloor L_{2} \rfloor\|_{F}^{2}] = c \frac{d_{1}(d_{1} - 1)}{2} \xi^{2} = \frac{d_{2}(d_{2} - 1)}{2} \xi^{2}.
        \]
        Then by assuming $\lfloor L_{1} \rfloor$ $\lfloor L_{2} \rfloor$ have priors independent of $\mathcal{D}(L_{1})$, $\mathcal{D}(L_{2})$, and letting
        \[
        M_{1} = \frac{d_{1} (d_{1} - 1)}{2}
        \]
        it then follows that
        \begin{align}
            \mathbb{E}[\| \lfloor L_{1} \otimes L_{2} \rfloor \|_{F}^{2} ] &= \mathbb{E} [\|\lfloor L_{1} \rfloor \|_{F}^{2}] \mathbb{E}[\|\mathcal{D}(L_{1}) \|_{F}^{2}] + \mathbb{E}[\| \mathcal{D}(L_{1})\|_{F}^{2}] \mathbb{E}[\| \lfloor L_{2} \rfloor \|_{F}^{2}] + \mathbb{E}[\|\lfloor L_{1} \rfloor \|_{F}^{2}] \mathbb{E}[\|\lfloor L_{2} \rfloor \|_{F}^{2}] \nonumber \\
            &\approx F_{D}^{\frac{1}{2}} (\mathbb{E} [\|\lfloor L_{1} \rfloor \|_{F}^{2}]  + \mathbb{E}[\| \lfloor L_{2} \rfloor \|_{F}^{2}]) + \mathbb{E}[\|\lfloor L_{1} \rfloor \|_{F}^{2}] \mathbb{E}[\|\lfloor L_{2} \rfloor \|_{F}^{2}] \label{eq: expected diagonal frob simplificiation} \\
            &= F_{D}^{\frac{1}{2}} M_{1}(1 + \frac{1}{c}) \xi + \frac{M_{1}^{2}}{c} \xi^{2} \label{eq: final quadratic form}
        \end{align}
        where (\ref{eq: expected diagonal frob simplificiation}) follows from Proposition \ref{prop: independent diagonal parameterization}. Further by assuming $\mathbb{E}[ \|\lfloor L_{1} \otimes L_{2} \rfloor \|_{F}^{2}] = F_{L}$, it then follows from (\ref{eq: final quadratic form}) that
        \[
        F_{D}^{\frac{1}{2}} M_{1}(1 + \frac{1}{c}) \xi + \frac{M_{1}^{2}}{c} \xi^{2} = F_{L}
        \]
        and solving the quadratic equation in $\xi$ yields (\ref{eq: beta variance}).
    \end{proof}
\end{theorem}

\begin{corollary}
    Suppose $\{L_{1}^{(i)}, L_{2}^{(i)}\}_{i = 1}^{K}$ be positive definite Cholesky factors such that $\{\mathbb{D}(L_{1}^{(i)}), \mathbb{D}(L_{2}^{(i)}) \} = \{D_{1}, D_{2} \}$ for all $i$, and let $L$ be the Cholesky factor defined by
    \[
    L = \sum_{i = 1}^{K} \lfloor L_{1}^{(i)} \otimes L_{2}^{(i)} \rfloor + D_{1} \otimes D_{2}
    \]
    with $D_{1}, D_{2}$ parameterized according to Proposition \ref{prop: independent diagonal parameterization}, and 
    \begin{align}
        \lfloor L_{1}^{(i)} \rfloor[i,j] &\sim N(0, \sigma^{2} = \omega_{i} \beta) \\
        \lfloor L_{2}^{(i)}\rfloor[p,q] &\sim N(0, \sigma^{2} = \omega_{i}\beta)
    \end{align}
    such that $\Omega = (\omega_{1}, \ldots, \omega_{K}) \in \Delta^{K}$ is distributed on the unit simplex and $\beta$ is from Theorem \ref{thm: lower triangular expected frobenius norm}, then
    \begin{equation}
        \mathbb{E}[\| \lfloor L \rfloor \|_{F}^{2} ] = F_{L}.
    \end{equation}
    \begin{proof}
       First note that
        \begin{align*}
        \mathbb{E}[\| \lfloor L_{1} \otimes L_{2}\|_{F}^{2}] = &\mathbb{E}[\| \lfloor \sum_{i} L_{1}^{(i)} \rfloor\|_{F}^{2}\|D_{2} \|_{F}^{2} + \|D_{1}\|_{F}^{2} \| \lfloor \sum_{i} L_{2}^{(i)} \rfloor \|_{F}^{2}] \\
        &+ \mathbb{E}[\|\lfloor \sum_{i} L_{1}^{(i)} \rfloor \|_{F}^{2}\|\lfloor \sum_{i} L_{2}^{(i)} \rfloor \|_{F}^{2}].
        \end{align*}
        Note also that if $\{A_{i}\}_{i = 1}^{K}$ are random matrices such that $A_{i} \stackrel{d}{=} \alpha_{i} A$ for independent random variables $\{\alpha_{i}\}_{i = 1}^{K}, A$:
        \[
         \sum_{i = 1}^{K} A_{i}  \stackrel{d}{=} \sum_{i = 1}^{K} \alpha_{i} A
        \]
        and therefore it follows that
        \[
        \| \sum_{i = 1}^{K} A_{i} \|_{F}^{2} \stackrel{d}{=} \|\sum_{i = 1}^{K} \alpha_{i} A \|_{F}^{2} = (\sum_{i = 1}^{K} \alpha_{i})^{2} \|A \|_{F}^{2}. 
        \]
        Let $\lfloor S_{j} \rfloor$ denote a $d_{j} \times d_{j}$ strictly lower triangular matrix with standard normal entries. Note $\| S_{j} \|_{F}^{2} \sim \chi_{d_{j}(d_{j - 1})/2}^{2}$ and $\|L_{1}^{(j)}\|_{F}^{2}\|L_{2}^{(j)}\|_{F}^{2} \stackrel{d}{=} (\omega_{i} \beta)^{2} Q$, where $Q \sim\|\lfloor S_{1} \rfloor\|_{F}^{2} \|\lfloor S_{2}\rfloor\|_{F}^{2}$ for independent $S_{1}, S_{2}$. From this, it is clear that
        \[
        \mathbb{E}[\| \lfloor \sum_{i} L_{j}^{(i)} \rfloor \|_{F}^{2} ] = \mathbb{E}[\mathbb{E}[\| \lfloor \sum_{i} L_{j}^{(i)} \rfloor \|_{F}^{2} \vert \Omega]] = \mathbb{E}[\mathbb{E}[(\sum_{i} \omega_{i} \sqrt{\beta})^{2} \| \lfloor S_{j} \rfloor \|_{F}^{2} \vert \Omega \|_{F}^{2} ]] = \beta \mathbb{E}[\| \lfloor S_{j} \rfloor \|_{F}^{2}].
        \]
        From Theorem \ref{thm: lower triangular expected frobenius norm}, it follows $\mathbb{E}[\|\lfloor L \rfloor \|_{F}^{2}] = \beta$.
    \end{proof}
\end{corollary}

\subsection{Non-Dynamic Bayesian Model Selection} \label{sec: Non-Dynamic}
Expressing the model formulation above more compactly, the full Sum of Cholesky factor Kronecker Product Decomposition (SCKPD) Bayesian model is specified as
\begin{equation} \label{eq: SCKPD Model}
\begin{aligned}
    y_{1},\ldots, y_{N} &\sim N(0, L^{\dagger} (L^{\dagger})^{T})\\
    L^{\dagger} \vert\{L_{1}^{(i)}, L_{2}^{(i)}\}_{i = 1}^{K}, D_{1}, D_{2} &= \sum_{i = 1}^{K} \lfloor L_{1}^{(i)} \otimes L_{2}^{(i)}\rfloor + D_{1} \otimes D_{2}\\
    D_{i} \vert a_{i}, \gamma_{D} &\sim \Gamma\big(a_{i}, \exp(\psi_{0}(a_{i}) - \frac{\gamma_{D}}{2 d_{1} d_{2}}) \big)\\
    a_{i} &= \arg \min_{\xi} \vert \xi^{2} + \xi - c_{i} \exp(2 \psi_{0}(\xi) \vert, \quad c_{i} = \frac{\sqrt{F_{D}}}{d_{i}} \exp(-\frac{\gamma_{D}}{2d_{1}d_{2}}) \\ 
    \lfloor L_{1}^{(i)} \rfloor \stackrel{D}{=} \lfloor L_{2}^{(i)} \rfloor \vert \Omega, \beta, F_{D}, F_{L}&\sim N(0, \sigma^{2} = \omega_{i} \beta),\\
   \text{where   }  \beta &= \frac{- M_{1} \sqrt{F_{D}} (1 + \frac{1}{c}) + \sqrt{F_{D}(1 + \frac{1}{c})^{2} + 4 (M_{1}^{2}/c)F_{L}}}{2M_{1}^{2}/c}\\
    \Omega\vert \theta &\sim \mathcal{D}(\theta) \\
    \theta &\sim U(0,1)
\end{aligned}
\end{equation}
where $\Omega = (\omega_{1}, \ldots, \omega_{k})$, $\gamma_{D} = \log \vert \mathcal{L}(S) \vert $ $F_{D} = \|\mathbb{D}(\mathcal{L}(S)) \|_{F}^{2}$, $F_{L} = \| \lfloor \mathcal{L}(S) \rfloor \|_{F}^{2}$ such that $S$ is the sample covariance matrix, $c = \frac{d_{1}(d_{1} - 1)}{d_{2} (d_{2} - 1)}$, $M = \frac{d_{1}(d_{1} - 1)}{2}$, and $\mathcal{D}(\theta)$ denotes the Dirichlet distribution with concentration parameter $\theta$.

\subsection{Seasonally Dynamic Covariance} \label{sec: seasonally dynamic covariance}
In this section, we will provide a simple extension of the SCKPD Bayesian model (\ref{eq: SCKPD Model}) in the case where we have collections of observations with seasonal matrix values that have fixed observations within the season, but the correlation structure shifts between seasons. More specifically, we consider the case where we have observations that are generated from a finite collection of Cholesky factors $C = \{L_{1}^{(i)}, L_{2}^{(i)}\}_{i = 1}^{K}$, where in season $s$ in cycle $c$, the relative contributions of the elements in $C$ to the precision matrix are constant for all $y_{1}^{(s,c)}, \ldots, y_{n}^{(s,c)}$, and change as we move to $y_{1}^{(s+1, c)}, \ldots, y_{n}^{(s+1, c)}$ or $y_{1}^{(s- S + 1, c + 1)}, \ldots, y_{n}^{(s - S +1, c + 1)}$. As we are considering the regime where the relative contributions shift but not the underlying covariances themselves, we provide a particularly simplistic form for the data generating process, which we term the Seasonally Dynamic sum of Cholesky factor Kronecker Product Decomposition (SD-SCKPD) Bayesian model:
\begin{equation} \label{eq: SD-SCKPD Model}
    \begin{aligned}
         y_{1}^{t},\ldots, y_{N}^{t} &\sim N(0, L_{t}^{\dagger} (L_{t}^{\dagger})^{T})\\
    L_{t}^{\dagger} \vert\{L_{1,t}^{(i)}, L_{2,t}^{(i)}\}_{i = 1}^{K}, D_{1}, D_{2} &= \sum_{i = 1}^{K} \lfloor L_{1,t}^{(i)} \otimes L_{2,t}^{(i)}\rfloor + D_{1} \otimes D_{2}\\
    D_{i} \vert a_{i}, \gamma_{D} &\sim \Gamma\big(a_{i}, \exp(\psi_{0}(a_{i}) - \frac{\gamma_{D}}{2 d_{1} d_{2}}) \big)\\
    a_{i} &= \arg \min_{\xi} \vert \xi^{2} + \xi - c_{i} \exp(2 \psi_{0}(\xi) \vert, \quad c_{i} = \frac{\sqrt{F_{D}}}{d_{i}} \exp(-\frac{\gamma_{D}}{2d_{1}d_{2}}) \\ 
    \lfloor L_{1,t}^{(i)} \rfloor \stackrel{D}{=} \lfloor L_{2,t}^{(i)} \rfloor \vert \Omega_{t}, \beta, F_{D}, F_{L}&\sim N(0, \sigma^{2} = \omega_{i,t} \beta),\\
   \text{where   }  \beta &= \frac{- M_{1} \sqrt{F_{D}} (1 + \frac{1}{c}) + \sqrt{F_{D}(1 + \frac{1}{c})^{2} + 4 (M_{1}^{2}/c)F_{L}}}{2M_{1}^{2}/c}\\
    \Omega_{1}\vert \theta &\sim \mathcal{D}(\theta) \\
    \Omega_{t + 1} &= A \Omega_{t}, \quad A \in \mathbb{R}_{+}^{K \times K}, \quad \sum_{j = 1}^{K} A_{i,j} = 1 \\
    \theta &\sim U(0,1).
    \end{aligned}
\end{equation}

In this case, $M_{1}, F_{D}, F_{L}, \gamma_{D}$, and $c$ maintain the same interpretation of the SCKPD model, but restricted to the first season of observations $y_{1}^{1}, \ldots, y_{N}^{1}$. This extension introduces the stochastic matrix of nonnegative columns $A$ to allow perturbations between $\Omega_{t}$ and $\Omega_{t + 1}$ while maintaining the simplex constraint:
\[
\sum_{k = 1}^{K} (A\Omega_{t + 1})_{k} = \sum_{i = 1}^{K} \sum_{j = 1}^{K} A_{i,j} \Omega_{i} = \sum_{i =1}^{K} \Omega_{i} = 1.
\]

Such a model can be useful for cases where data may exhibit varying degrees of deeply correlated residual structure to varying degrees during different seasons. As we emphasize the use of this model in modeling seasonality effects, we can equivalently state that for $\Omega(0) = \Omega$, it follows that at season $s$ in cycle $c$:
    \[
    \Omega(s,c) = A^{c + s -1} \Omega
    \]

Note that such a prior assumes that the covariance exhibits a temporal relationship in $\Sigma_{t} = L_{t}^{\dagger} \big[ L_{t}^{\dagger} \big]^{T}$ such that

\begin{align*}
    \mathbb{E}[tr(\Theta_{t})] &= \mathbb{E}\big[\sum_{i} (L_{ii}^{\dagger})^{2} +  \sum_{i} \sum_{j > i} (L_{i,j}^{\dagger})^{2} \big] = \mathbb{E}[\| \mathbb{D}(L^{\dagger})\|_{F}^{2}] + \mathbb{E}\big[\| \lfloor L^{\dagger}\rfloor \|_{F}^{2}] = F_{D} + F_{L}\\
    \mathbb{E}\big[ \log \vert \Theta_{t} \vert \big] &= \gamma_{D}
\end{align*}
Implicates a temporal evolution where the total variance and eigenvalues' geometric mean is fixed across time states. This however is only a restriction on the magnitudes of variation, rather than the directions, allowing for flexibility in modeling directions of regime shifts in cases where a time series may effectively exhibit patterns where variability of a vector time series is "passed" between it's elements through a temporal evolution.
\section{Implementation Details} \label{sec: Implementation details}
We sample the SCKPD model using Hamiltonian Monte Carlo. However, this poses two options: to convert to an unconstrained parameterization for sampling, as is done with {\tt stan}, or to use a geometrically informed implementation such as geodesic Monte Carlo (\cite{holbrook2018geodesic, simonis2025separablegeodesiclagrangianmonte, byrne2013geodesic}). Under the log-Cholesky metric, the geodesic connecting $L_{0} = \mathcal{L}(P_{0})$ to $L_{1} = \mathcal{L}(P_{1})$ is given by \cite{lin2019riemannian}:
\begin{align*}
    &\lfloor L_{0} \rfloor + t(\lfloor L_{1} \rfloor - \lfloor L_{0} \rfloor) + \mathbb{D}(L_{0}) \exp (t \mathbb{D}(L_{0}) \log (\mathbb{D}(L_{0})^{-1} \mathbb{D}(L_{1})) \mathbb{D}(L_{0})^{-1}) \\
    &= \lfloor L_{0} \rfloor + t(\lfloor L_{1} \rfloor - \lfloor L_{0} \rfloor) + \mathbb{D}(L_{1})^{t} \mathbb{D}(L_{0})^{1-t}.
\end{align*}

A consideration in using a geodesic implementation is the potential efficiency gained by the Riemannian analogue of Euclidean convexity, as was discovered for multiway covariances of Gaussian arrays in \cite{wiesel2012convexity}.
\begin{definition}
    Let $(\mathcal{M},g)$ be a Riemannian manifold. A function $f$ defined on $\mathcal{M}$ is geodesically convex if for all $p,q \in \mathcal{M}$:
    \[
    f(\gamma(t)) \leq t f(\gamma(0)) + (1- t) f(\gamma(1))
    \]
    where $\gamma(t)$ is the geodesic such that $\gamma(0) = p$, $\gamma(1) = q$.
\end{definition}
However, as an extension of Lemma 3 of (\cite{wiesel2012convexity}), we show that the inner product for vectorized tensor normal observations is not jointly geodesically convex under a product manifold geometry for the log-Cholesky metric.
\begin{lemma} \label{lemma: geodesically non-convex cholesky}
    Let $\mathbb{C}_{n} = \mathcal{C}_{1} \times \mathcal{C}_{2} \times \cdots \times \mathcal{C}_{n}$ denote the product manifold of $\{ \mathcal{C}_{i} \}_{i = 1}^{n}$, independently endowed with the Cholesky metric. That is, each element is the n-tuple pair, $\mathcal{M} = (\times_{i = 1}^{n} C_{i}, \oplus G(C_{i}))$. Suppose $h \in \mathbb{R}^{n}$, then $h^{T} LL^{T} h$ is not geodesically convex in $\times_{i = 1}^{D} L_{i} \in \mathcal{M}$.
    \begin{proof}
Suppose $L_{1} \in \mathcal{C}^{n}, L_{2} \in \mathcal{C}^{m}$. For the Kronecker product of Cholesky factors,
\begin{align}
    &h^{T} (L_{1}(t) \otimes L_{2}(t)) (L_{1}(t) \otimes L_{2}(t))^{T}  h = \sum_{i} y_{i}(t) \\
    &y_{i}(t) = \sum_{j = 1}^{i} [(L_{1}(t) \otimes L_{2}(t))_{i,j} h_{j}]^{2}.
\end{align}
Noting that 
\[
(L_{1}(t) \otimes L_{2}(t))_{mj + s, mq + t} = (L_{1}(t))_{j,q}(L_{2}(t))_{s,t}, \quad 1 \leq j,q \leq n - 1, \quad 1\leq s,t \leq m.
\]
Results in:
\begin{equation} \label{eq: y equation cholesky}
    y_{mj+s}(t) = \sum_{0 \leq (a,b) \leq (j,s)} ([L_{1}(t)]_{j,a}^{2} [L_{2}(t)]_{s,b}^{2} h_{ma + b}^{2})
\end{equation}
where 
\begin{align*}
[L_{i}(t)]_{x,y} &= \begin{cases}
    (1-t) \lfloor L_{i}(0) \rfloor_{x,y} + t \lfloor L_{i}(1)\rfloor_{x,y} \text{ if } x \neq y \\
    \mathbb{D}_{x}(L_{i}(0))^{1-t} \mathbb{D}_{x}(L_{i}(1))^{t} \text{ if } x = y
\end{cases} \\[1.5 em]
\implies [L_{i}(t)]_{x,y}^{2} &= \begin{cases}
    (1-t)^{2} \lfloor L_{i}(0) \rfloor_{x,y}^{2} + t^{2} \lfloor L_{i}(1)\rfloor_{x,y}^{2} + 2(1-t)t\lfloor L_{i}(0) \rfloor \lfloor L_{i}(1) \rfloor_{x,y} \text{ if } x \neq y \\
    \mathbb{D}_{x}(L_{i}(0))^{2(1-t)} \mathbb{D}_{x}(L_{i}(1))^{2t} \text{ if } x = y.
    \end{cases}
\end{align*}

Then any element of $y$ gives no guarantee of convexity, as we cannot guarantee positivity, nor strictly increasing/decreasing in any of the entries due to any second-order or higher terms containing products of unconstrained lower triangular elements. Hence, the function is certainly not jointly geodesically convex in this space.
    \end{proof}
\end{lemma}

Practically, Lemma \ref{lemma: geodesically non-convex cholesky} states that if we follow Riemannian gradients under the product manifold geometry of Cholesky factors for just a single Kronecker product alone, it is not a geodesically convex path. One could potentially use parallel tempering as an attempt to alleviate the non-convexity. However, this would not be of great benefit when off the shelf Hamiltonian Monte Carlo samplers such as {\tt stan} exist with built-in tuning mechanisms for efficiency.

    For implementation with {\tt stan}, gradients are instead computed with respect to unconstrained parameterizations of $L_{i}^{(j)}$. We need not deal directly with the conversion of constrained lower triangular matrices to an unconstrained space, which is handled internally within {\tt stan}. However, it would be important to consider alternative forms of the likelihood for efficiency. In Section \ref{sec: Computational results}, we cover implementation details for efficient computation and parallelization of {\tt stan}'s computing environment for Bayesian SCKPD model.

    However, one note is that the estimation of $A$ for the dynamic covariance model of Section \ref{sec: seasonally dynamic covariance}: {\tt stan} does not allow directly placing priors on the columns of a matrix parameter. In each example considered within this paper, we assume that the columns are iid distributed according to a $\mathcal{D}(\alpha\mathbf{1}_{k})$. By \cite{devroye2006nonuniform}, we can instead treat $A$ as a transformed parameter:
    \begin{align*}
        A[i,j] \vert A^{*} &= \frac{A^{*}[i,j]}{\sum_{j} A^{*}[i,j]} \\
        A_{i,j} &\stackrel{iid}{\sim} \Gamma(\alpha,1)
    \end{align*}

    \section{Simulated Data Examples} \label{sec: Simulated Data Examples}
    In this section we will consider simulated data examples to assess our model's ability to recover the true data generating process.
    \subsection{Static Covariance} \label{subsec: static covariance simulated data example}
\begin{figure}[h!]
    \centering
    % First row (2 images)
    \includegraphics[width=0.45\textwidth]{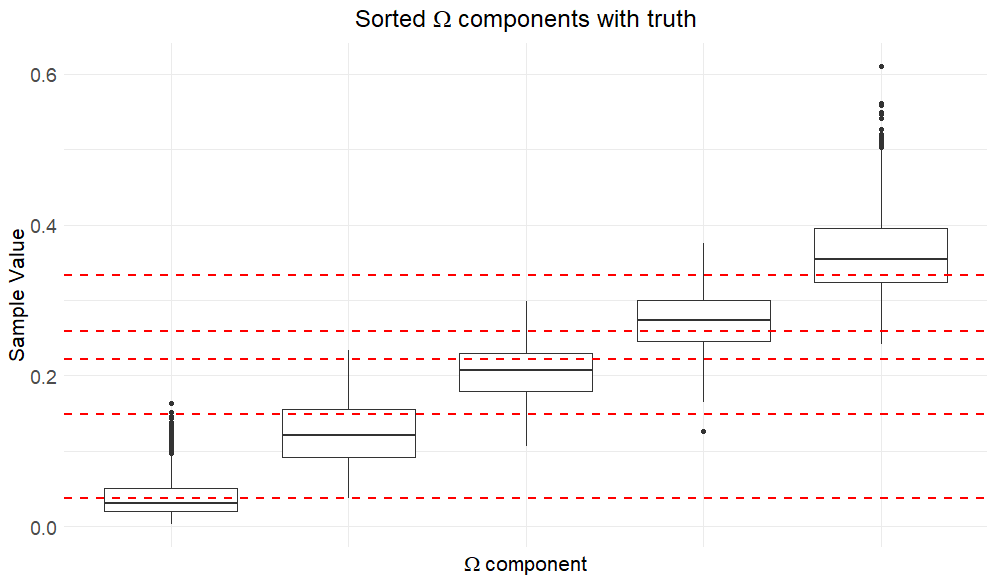}
    \hfill
    \includegraphics[width=0.45\textwidth]{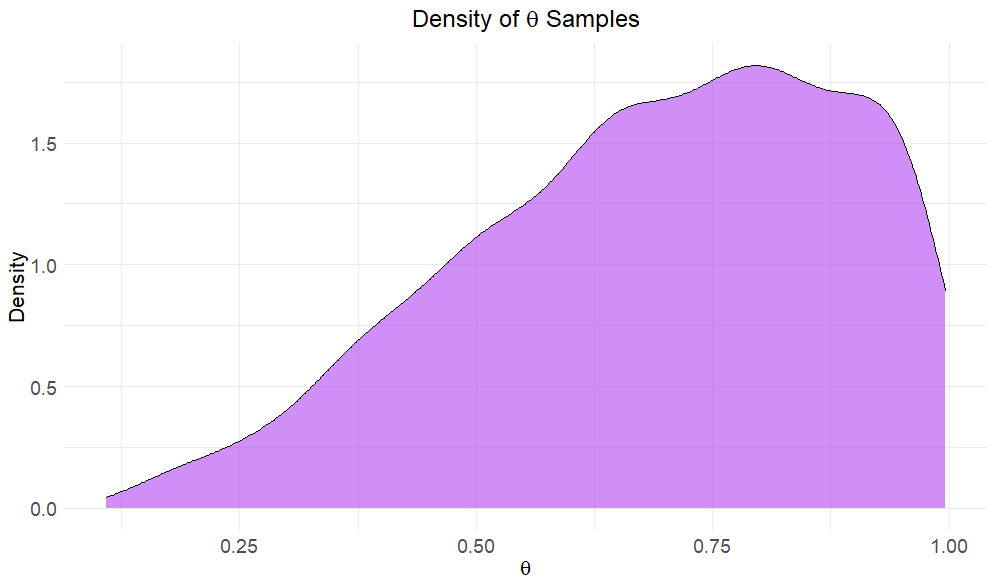}

    \vskip\baselineskip  % Space between rows

    % Second row (3 images)
    \includegraphics[width=0.3\textwidth]{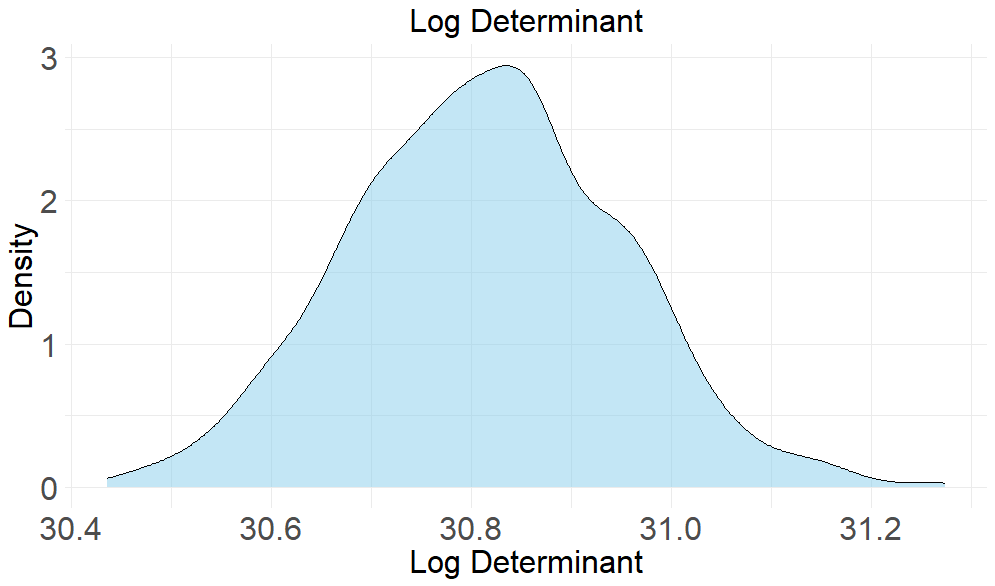}
    \hfill
    \includegraphics[width=0.3\textwidth]{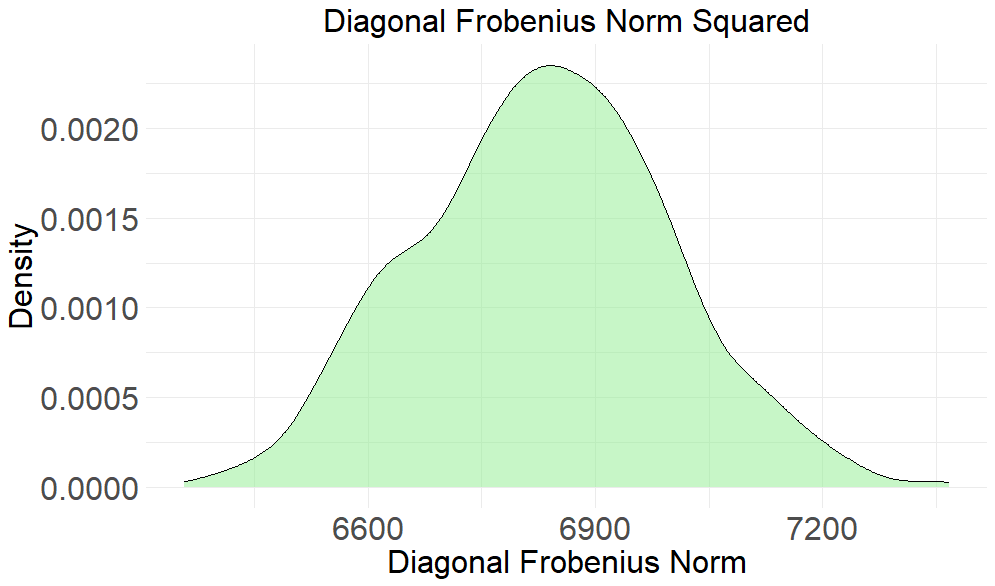}
    \hfill
    \includegraphics[width=0.3\textwidth]{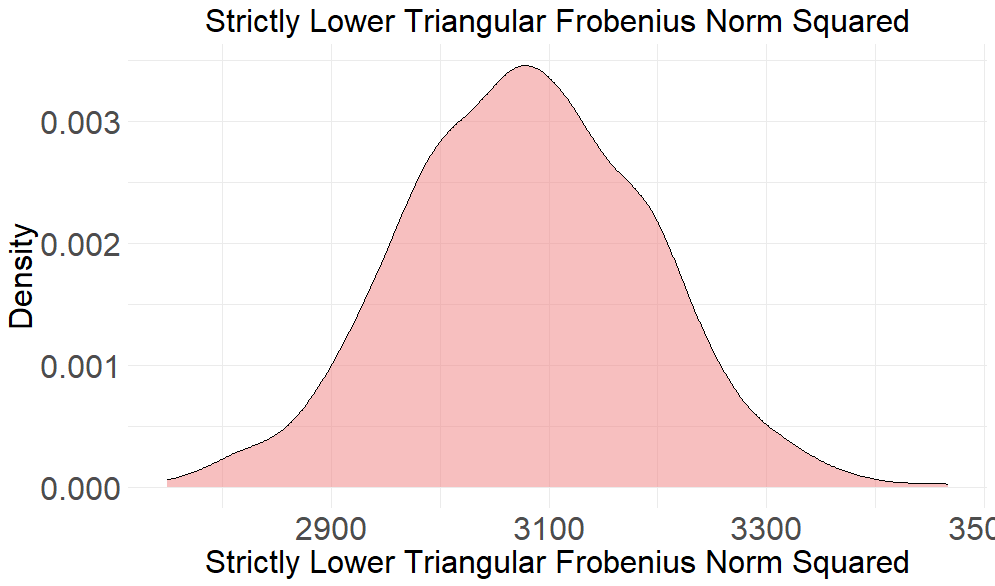}

    \caption{Posterior samples of the SCKPD model with $k = 5, \text{ } r = 5$ with $d_{1} = 4, \text{ } d_{2} = 5$. Top row: $\Omega$ components (left) with ground truth in dotted red line and $\theta$ (right). Bottom row: $\log \vert L^{\dagger} \vert $ (left), $\|\mathbb{D}(L^{\dagger}) \|_{F}^{2}$ (middle), $\| \lfloor L^{\dagger} \rfloor \|_{F}^{2}$ (right). Note the strong clustering of both the posterior samples of $\Omega$ and concentration of $\theta$ near 1. }
     \label{fig: data generating process 1}
\end{figure}

    In our first simulation, we consider the fixed mean case where the true data generating process is simulated according to:
    \begin{align*}
        y &\sim N(0, \Theta = L^{\dagger} [L^{\dagger}]^{T})\\
        L^{\dagger} &= \sum_{j = 1}^{K} \lfloor L_{1}^{(i)} \otimes L_{2}^{(i)}\rfloor + D_{1} \otimes D_{2}\\
        \lfloor L_{1}^{(i)} \rfloor \stackrel{d}{=} \lfloor L_{2}^{(i)}\rfloor &\sim N(0, \omega_{i} \beta)
    \end{align*}

    with $\beta = 2$, $\Omega = \frac{\Omega_{u}}{\sum_{j} \Omega_{u}[j]}$ where $\Omega_{u} = (1,4,6,7,9)$. In the first 3 simulation examples, we generate data and fit the SCKPD Bayesian model according to:
    \begin{itemize}
        \item $r = k = 5$ (perfect association of true separability rank and data generating separability rank)
        \item $r = 5$, $k = 8$ (overspecification of separability rank)
        \item $r = 1$, $k = 5$ (overspecification of separability rank under a truly separable data generating process)
    \end{itemize}
    In all three examples, $dim(A) = 5$, $dim(B) = 4$, $n = 500$, and $r = 5$. In each example, strictly lower triangular entries were generated according to the prior distribution, and diagonals were generated by extracting diagonals from \begin{align}
        D_{1} &\sim diag(W(d_{1} + 2, \mathcal{D}_{1})), \quad D_{2} \sim diag(W(d_{2} + 2, \mathcal{D}_{2})) \\
        \mathcal{D}_{1} &= diag(.75,1,.2,.3,.1), \quad \mathcal{D}_{2} = diag(1,.4,.3,.2)
    \end{align}
    According to the SCKPD Bayesian model, our first simulation example is performed under choice $k = 5$. In all the examples considered, $a_{i}^{2} + a_{i} - \frac{\sqrt{F_{D}}}{d_{i}}\exp(-\gamma_{D}/(2d_{1}d_{2})) = \epsilon$ was solved using the {\tt nleqslv} package \cite{hasselman2018package} in R, and each model was run using the supplemental computational results using {\tt stan}'s reduce sum functionality for parallelization across the $r^{2}$ P-VL summands. The resulting posterior summarizations are found in Figure \ref{fig: data generating process 1}.

\begin{figure}[h!]
    \centering
    % First row (2 images)
    \includegraphics[width=0.45\textwidth]{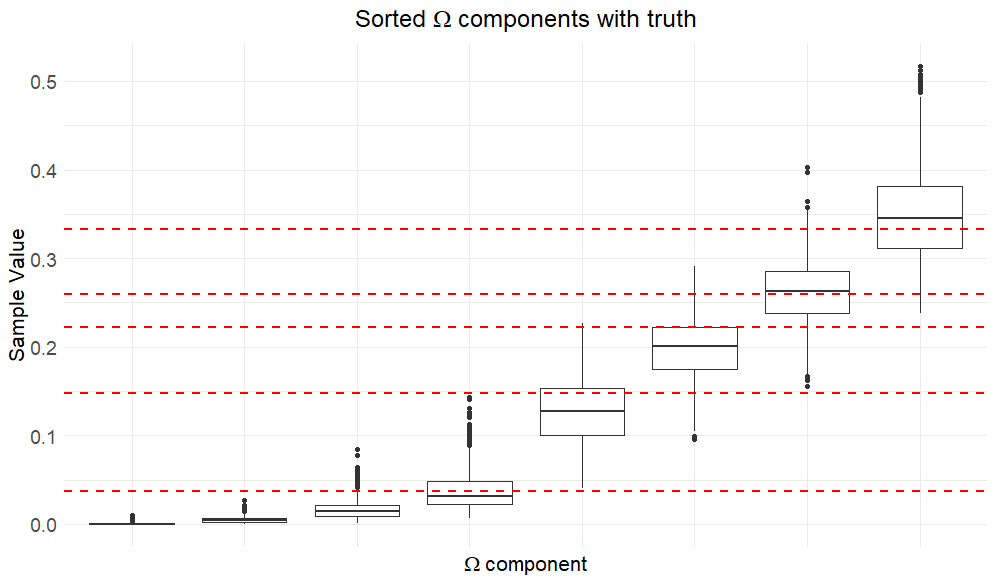}
    \hfill
    \includegraphics[width=0.45\textwidth]{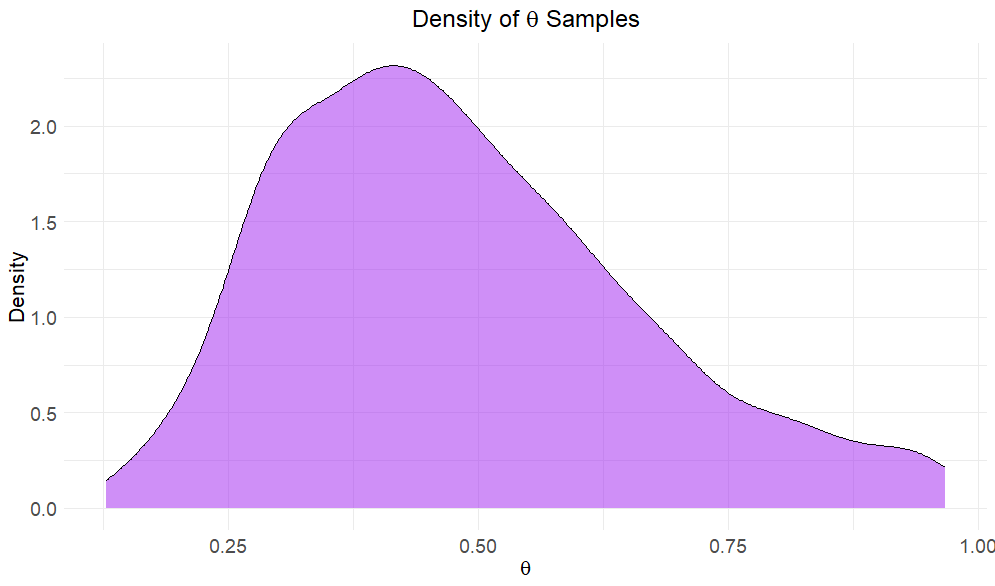}

    \vskip\baselineskip  % Space between rows

    % Second row (3 images)
    \includegraphics[width=0.3\textwidth]{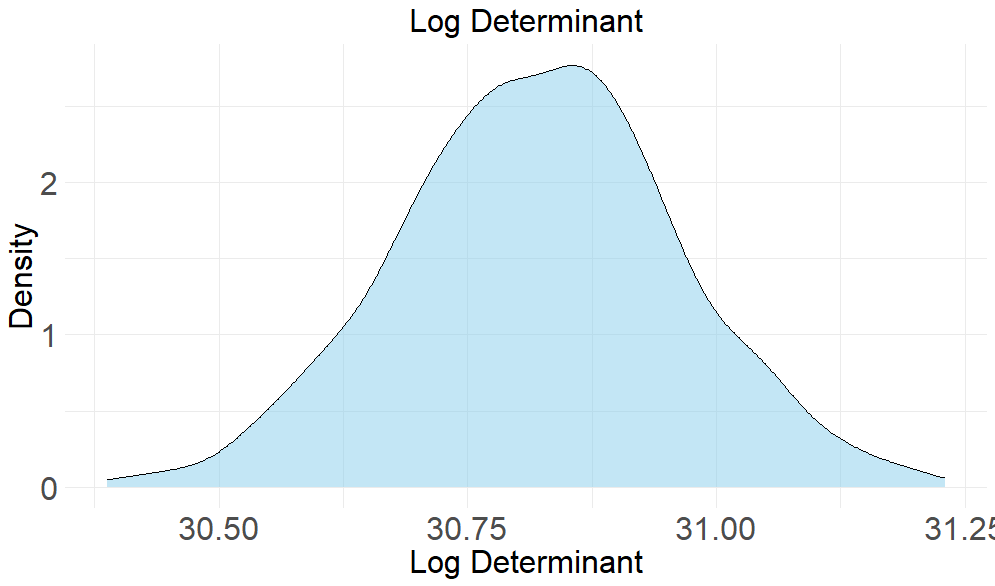}
    \hfill
    \includegraphics[width=0.3\textwidth]{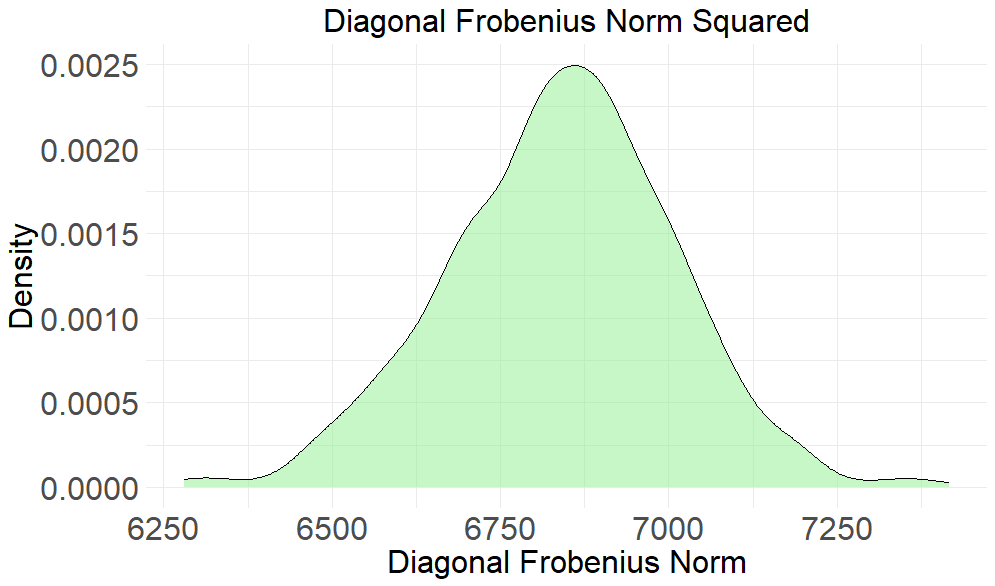}
    \hfill
    \includegraphics[width=0.3\textwidth]{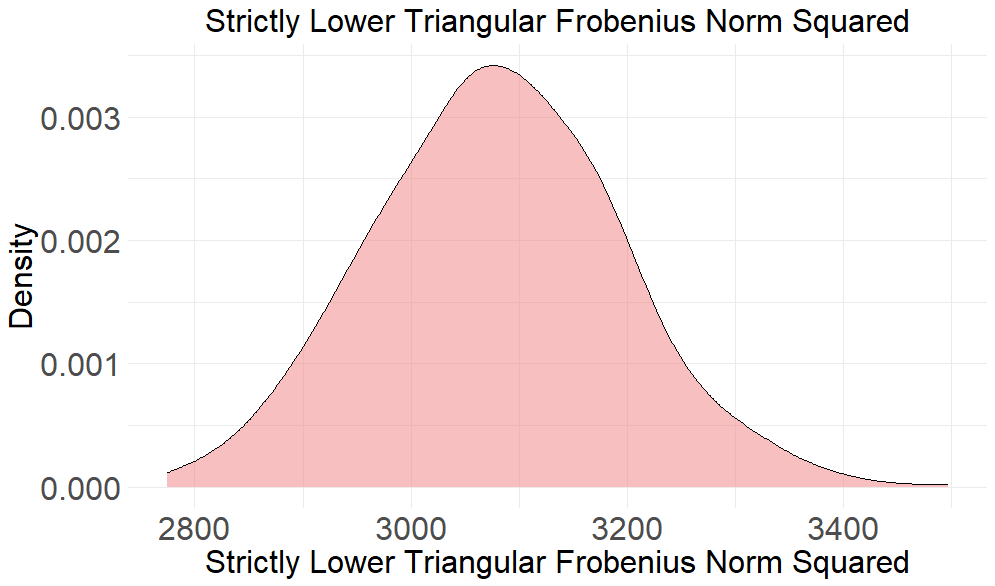}

    \caption{Posterior samples of the SCKPD model with $k = 8, \text{ } r = 5$ with $d_{1} = 4, \text{ } d_{2} = 5$. Top row: $\Omega$ components (left) with ground truth in dotted red line and $\theta$ (right). Bottom row: $\log \vert L^{\dagger} \vert $ (left), $\|\mathbb{D}(L^{\dagger}) \|_{F}^{2}$ (middle), $\| \lfloor L^{\dagger} \rfloor \|_{F}^{2}$ (right). Note the strong clustering of both the posterior samples of $\Omega$ and concentration of $\theta$ near 1. }
     \label{fig: data generating process 1 part 2}
\end{figure}

In our second example, we modify the first data generation process to $r = 1$, leaving $k = 5$ to demonstrate the robust performance of the model to over-specification. The results of this posterior analysis are found in Figure \ref{fig: data generating process 1 part 2}.

\begin{figure}[h!]
    \centering
    \includegraphics[width=0.3\textwidth]{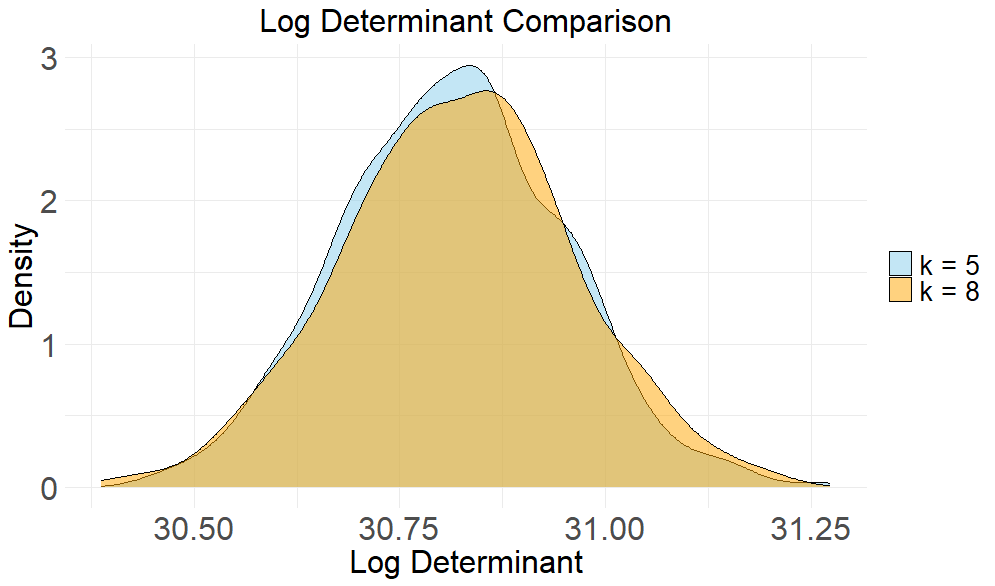}
    \hfill
    \includegraphics[width=0.3\textwidth]{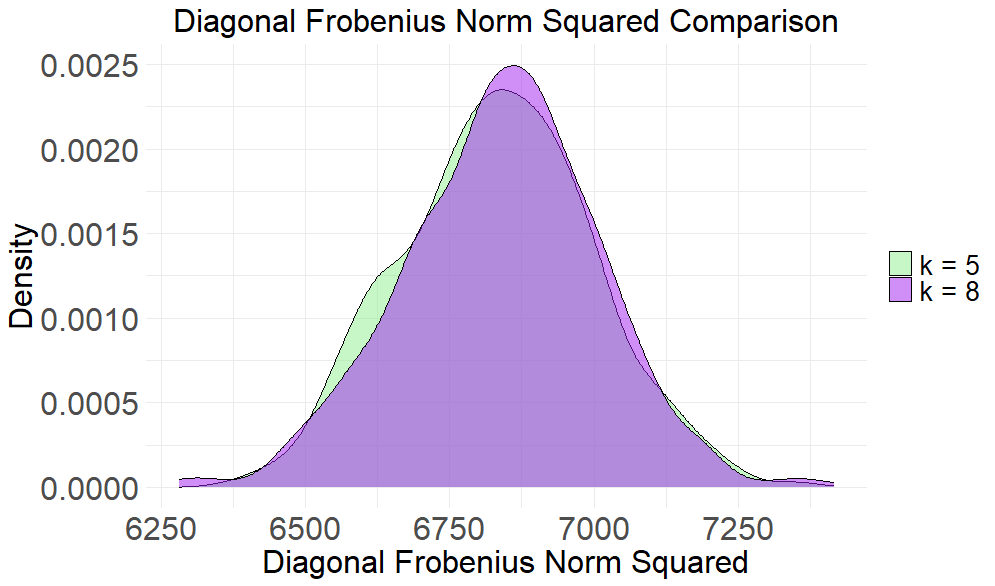}
    \hfill
    \includegraphics[width=0.3\textwidth]{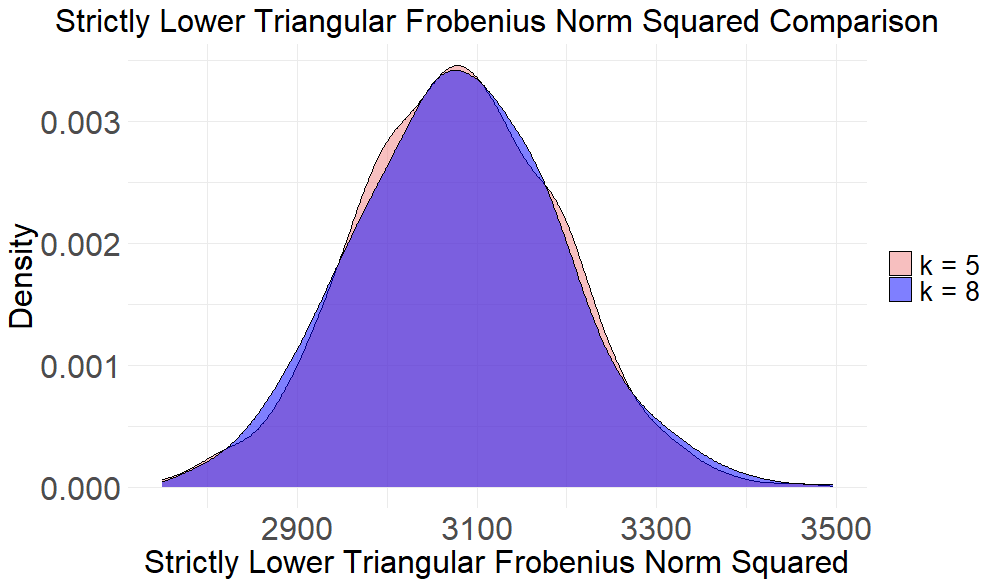}

    \caption{Comparison of posterior samples of the SCKPD model with $k = 5$ vs $k = 8$, and $\text{ } r = 5$. From left to right: $\log \vert L^{\dagger} \vert$, $\| \mathbb{D}(L^{\dagger}) \|_{F}^{2}$, and $\| \lfloor L^{\dagger} \rfloor \|_{F}^{2}$.}
     \label{fig: Posterior comparison}
\end{figure}

We note that the global matrix statistics are robust to exact or over-specification due to the centering of the priors, as demonstrated in Figure \ref{fig: Posterior comparison}.

\begin{figure}[h!]
    \centering
    % First row (2 images)
    \includegraphics[width=0.45\textwidth]{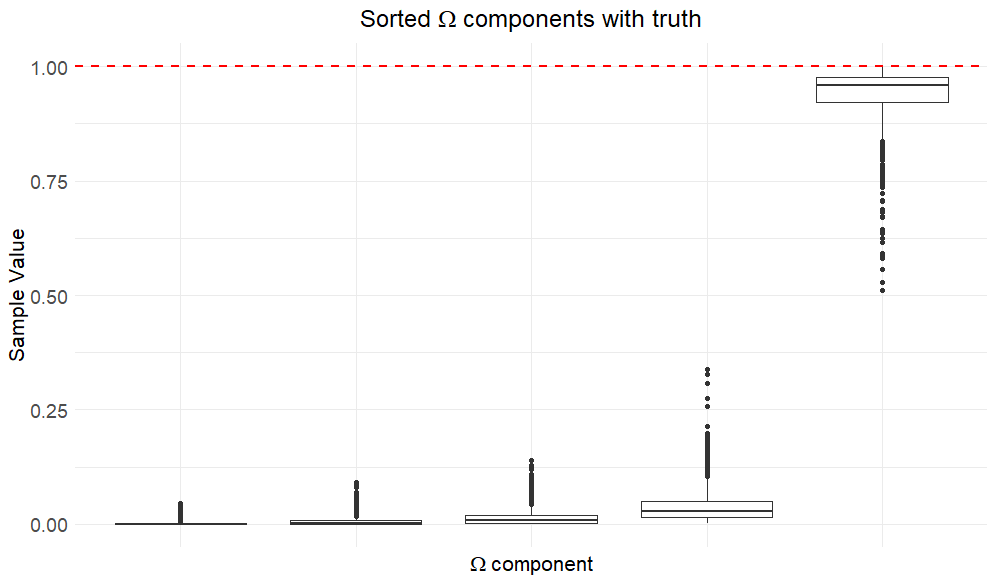}
    \hfill
    \includegraphics[width=0.45\textwidth]{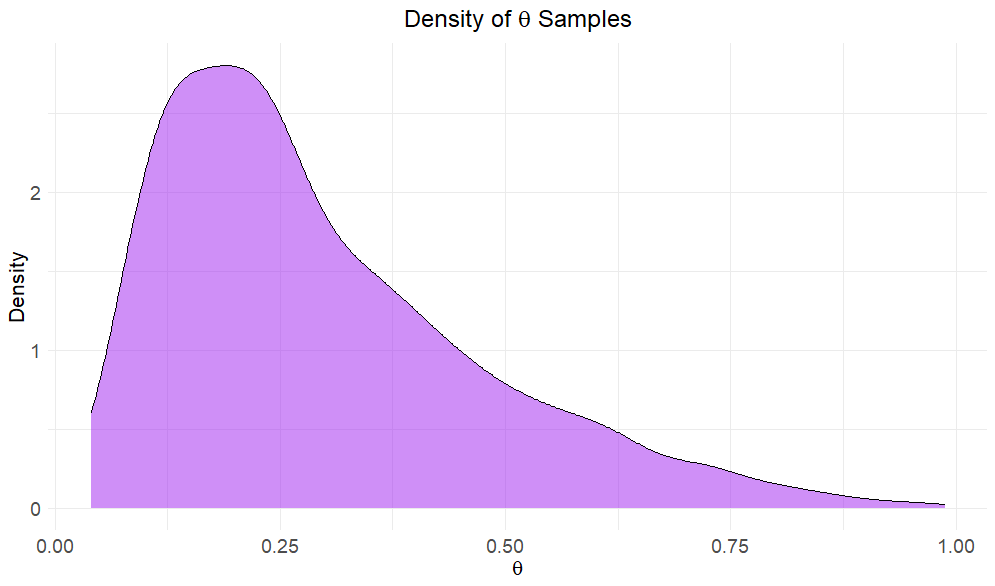}

    \caption{Posterior samples of the SCKPD model with $k = 5, \text{ } r = 1$ with $d_{1} = 4, \text{ } d_{2} = 5$. $\Omega$ components (left) with ground truth in dotted red line and $\theta$ (right). Note the strong clustering of both the posterior samples of $\Omega_{(5)}$ near 1 and concentration of $\theta$ near 0. }
     \label{fig: data generating process 2}
\end{figure}

Our third analysis focuses on the case of identifying true separability (i.e. the data generating process is focused on the case $R = 1$. In this case, we maintain the assumption $K = 5$ for overspecification, and evaluate the model's ability to recover only 1 non-zero component in $\Omega$. The results of this analysis are shown in Figure \ref{fig: data generating process 2}.

In our third example, we modify the first data-generating process to $r = 1$, leaving $k = 5$ to illustrate the model's ability to detect true separability. We do not show global matrix statistics in this figure, as we are not comparing posterior consistency between different configurations in this setting, but rather a different data generating process. These results are found in Figure \ref{fig: data generating process 2}.

\subsection{Seasonally Dynamic Covariance} \label{subsec: seasonally dynamic covariance simulated examples}
     In this section we will apply the SD-SCKPD model of Section (\ref{sec: seasonally dynamic covariance}) to an example where seasonal correlations are generated from a finite number of $K$ Cholesky factors:
    \begin{align*}
        \Theta(t) &= L_{t}^{\dagger} [L_{t}^{\dagger}]^{T}\\
        \lfloor L_{1}^{(i)} \rfloor \stackrel{D}{=} \lfloor L_{2}^{(i)} \rfloor \vert \Omega, \beta, F_{D}, F_{L}&\sim N(0, \sigma^{2} = \omega_{i}(t) \beta)\\
        \Omega(t + 1) &= A\Omega(t)
    \end{align*}
    where $A \in \mathbb{R}_{+}^{K \times K}$ is a non-negative column-stochastic matrix:
    \begin{align*}
        &A_{i,j} \geq 0 \quad \forall i,j\\
        &\sum_{j = 1}^{K} A_{i,j} = 1 \quad \forall i.
    \end{align*}
    
\begin{figure}[h!]
        \centering
        \includegraphics[width=0.95\linewidth]{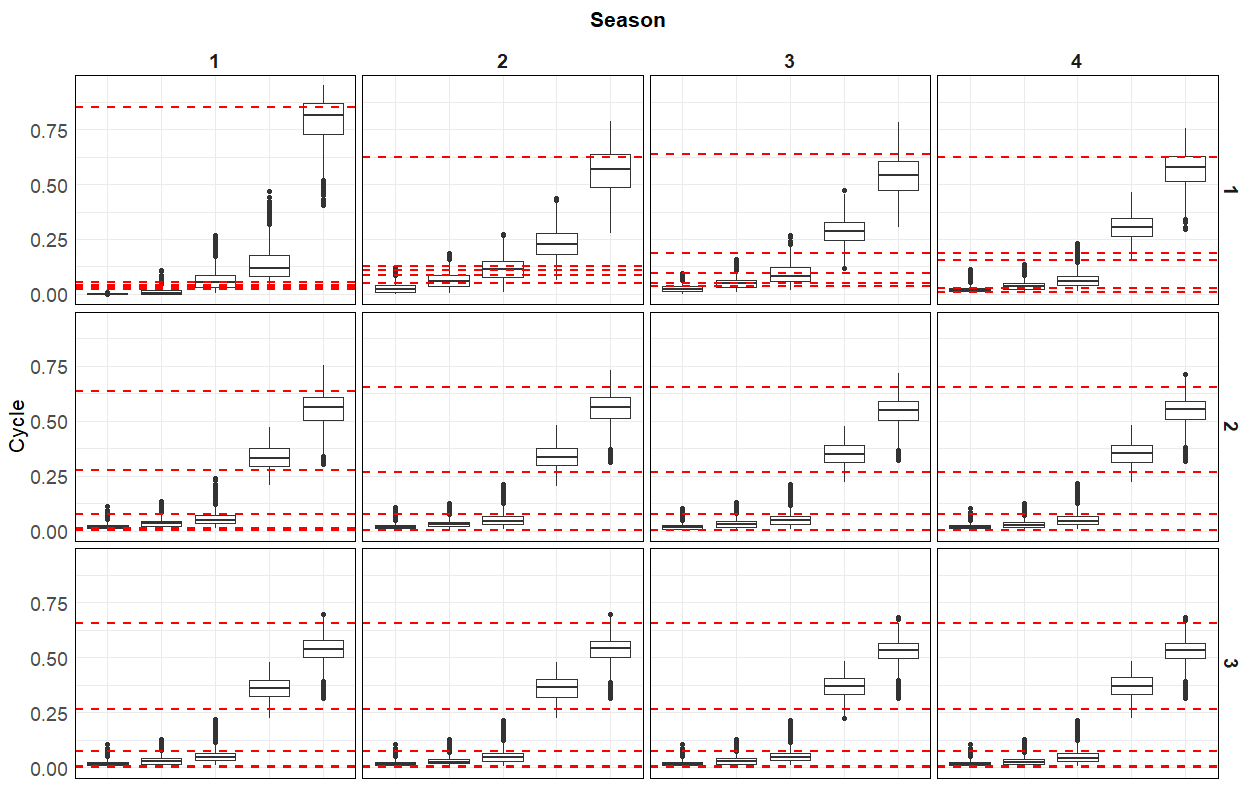}
        \caption{Ordered posterior samples of $\Omega(c,s)$ for seasonally changing Covariance. Seasonal changes are ordered across columns, cycle changes across rows. Horizontal red lines signify ground truth $\Omega_{c,s}$ at cycle $c$, season $s$, $c \in \{1,2,3\}$, $s \in \{1,2,3,4\}$. Note the model can correctly detect regime shifts within a $95\%$ credible interval while maintaining an appropriate sparsity level of $\Omega_{(1)}(c,s)$ and $\Omega_{(2)}(c,s)$. Moreover, the posterior adequately reflects the convergence of the steady-state distribution of the true $\Omega(c,s)$.}
        \label{fig:Diverse Omega}
    \end{figure}

    In this simulation example, we chose $d_{1} = 5$, $d_{2} = 2$, with $k = 5$, $r = 5$, with $n = 500$ per seasonal setting per cycle, and the columns of $A$ were simulated according to 
    \[
    A[,i] \sim \mathcal{D}(.05).
    \]
    The results of the dynamic covariance model can be found in Figure \ref{fig:Diverse Omega}.

    \section{Real Data Examples} \label{sec: real data examples}
    \subsection{Wisconsin Breast Cancer} \label{subsec: Wiscosnin Breast Cancer}
    In this section we provide a Bayesian analysis for the degree of data precision matrix separability for the Wisconsin Breast Cancer dataset \cite{agarap2018breast} using the model from Section \ref{sec: Non-Dynamic}. In particular, the Wisconsin Breast Cancer dataset is comprised of measurements of breast cancer cell nuclei derived from images using fine needle aspiration of breast mass. Statistics of $\{${\tt Mean, Standard Error, Worst} $\}$ are computed for the features $\{$ {\tt Radius, Texture, Perimeter, Area, Smoothness, Compactness, Concavity, Number of Concave Points, Symmetry, Fractal Dimension} $\}$. For our analysis, we focus on the subset of characteristic statistics $\{$ {\tt Mean, Worst} $\}$, and features $\{$ {\tt Radius, texture, concavity, symmetry, fractal dimension} $\}$ of malignant tumors. In total, this yields 212 matrix observations $\mathbf{Y}_{1}, \ldots, \mathbf{Y}_{212} \in \mathbb{R}^{2\times 5}$. 
    
\begin{figure}[h!]
        \centering
        \includegraphics[width=0.65\linewidth]{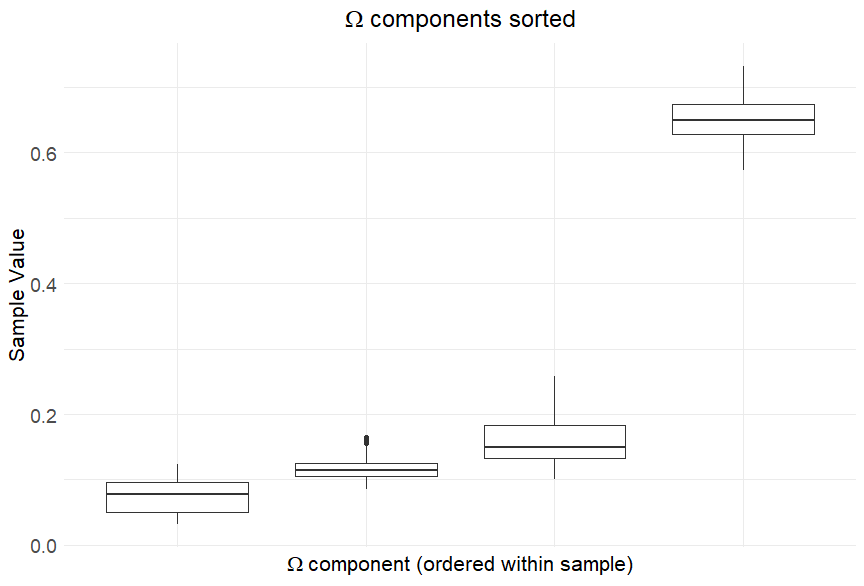}
        \caption{Sorted posterior $\Omega$ components under the SCKPD-B model for the analysis of the Wisconsin breast cancer dataset. Note all 4 components are non-zero, indicating a lack of separability in the precision matrix.}
        \label{fig: Wisconsin Breast Cancer Omega}
    \end{figure}

    We assume the SCKPD Bayesian model for the precision matrix:
    \begin{align*}
        vec(\mathbf{Y}_{1}), \ldots, vec(\mathbf{Y}_{567}) &\sim N(\bar{\mathbf{Y}}, \Theta =  L^{\dagger} [L^{\dagger}]^{T}) \\
        L^{\dagger} \vert \{L_{1}^{(i)}, L_{2}^{(i)}\}, \Omega  &\sim SCKPD( \{L_{1}^{(i)}, L_{2}^{(i)}\}, \Omega), \quad L_{1}^{(i)} \in \mathcal{L}^{+}(5), \text{ }L_{2}^{(i)} \in \mathcal{L}^{+}(2).
    \end{align*}
    If separability were true, by P-VL, then the degree of separability is upper bounded by $4$, which we assume for $K$ in our analysis. The results of this analysis are found in Figure \ref{fig: Wisconsin Breast Cancer Omega}.

    \subsection{Continental United States Climate Analysis} \label{sec: Climate Analysis}
     In this section we apply the results of Section \ref{sec: seasonally dynamic covariance} to a Bayesian regression analysis of climate data in the continental United States. The data set consists of 48 cities in the continental United States, and our analysis focuses on the multivariate observations {\tt TMAX, TMIN}, which are respectively the maximum and minimum ground temperature observations in tenths of degrees Celsius collected between spring 2020 and winter 2023.
    
     We focus our analysis on inference of the covariance of the coefficients corresponding to periodic trends across the different cities. Specifically, we fit the following univariate OLS model independently to each of the 48 cities where the output is $\hat{Y}_{c,t,1} = \hat{T}_{c,t,max}$, $\hat{Y}_{c,t,2} = \hat{T}_{c,t,min}$. Specifically:
     \begin{equation} \label{eq: Independent OLS models}
         \hat{Y}_{c,t,i} = \alpha_{c,i} + \sum_{j = 1}^{3} \big(\phi_{j,c,i} Y_{c,t-j,1} + \xi_{j,c,i} Y_{c,t - j, 2}\big) + \gamma_{1,c,i} \sin\big(2 \pi \frac{\Xi(t)}{365} \big) + \gamma_{2,c,i} \cos \big( 2 \pi \frac{\Xi (t)}{365} \big) + \epsilon
\end{equation}
that is, we fit OLS models on a city-dependent intercept, AR(3) components with lags on $\hat{T}_{c,t,max}$ and $\hat{T}_{c,t,min}$, and annual periodic patterns, where $\frac{\Xi(t)}{365}$ denotes the faction of time that has occurred within the season at that point. For example, assuming exactly $365$ observations occurred for each year, $\frac{\Xi(500)}{365} = \frac{500 - 365}{365} = \frac{135}{365}$, indicating that observation 500 occurs at $\frac{135}{365}$ the first part of year 2. 

     For these separate OLS models, we achieved a mean R-squared of $.884$ and $.831$ and a mean standard error of $28.8$ and $37.8$ for tenths of degree Celsius for $T_{min}$ and $T_{max}$, respectively. Table (\ref{table: significance proportions}) below illustrates the proportion of models that were significant $\big( \mathbb{P}(\theta > \vert t\vert) < .05)$. 
     
     \begin{table}     
     \centering
     \begin{minipage}{\linewidth}
         \centering
         \begin{tabular}{|c|c|c|c|c|c|c|c|c|}
         \hline
         $\alpha_{\cdot, 1}$ & $\gamma_{1, \cdot, 1}$ & $\gamma_{2, \cdot, 1}$ & $\phi_{1, \cdot, 1}$ & $\phi_{2, \cdot, 1}$ &  $\phi_{3, \cdot, 1}$ & $\xi_{1, \cdot, 1}$ & $\xi_{2, \cdot, 1}$ & $\xi_{3, \cdot, 1}$ \\ \hline
          $.47$  &
           $1$ &
           $1$ &
          $1$ &
            $.55$ &
           $.62$ &
           $1.00$ &
           $.85$ &
          $.13$  \\
          \hline
     \end{tabular}
     \end{minipage}
          \rule{0pt}{12pt} 
         \begin{minipage}{\linewidth}
         \centering
         \begin{tabular}{|c|c|c|c|c|c|c|c|c|}
         \hline
         $\alpha_{\cdot, 2}$ & $\gamma_{1, \cdot, 2}$ & $\gamma_{2, \cdot, 2}$ & $\phi_{1, \cdot, 2}$ & $\phi_{2, \cdot, 2}$ &  $\phi_{3, \cdot, 2}$ & $\xi_{1, \cdot, 2}$ & $\xi_{2, \cdot, 2}$ & $\xi_{3, \cdot, 2}$ \\ \hline
          $1$ &
           $1$ &
           $1$ &
           $.64$ &
            $.7$ &
            $.17$ &
           $1$ &
          $.68$ &
           $.49$ \\ \hline
     \end{tabular}
         \end{minipage}        
     \caption{Proportion of significance for OLS fits of equation (\ref{eq: Independent OLS models}) across the 48 datasets for TMIN (top row) and TMAX (bottom row). Note the significance of coefficients for annual cyclic patterns for both variables.}
     \label{table: significance proportions}
     \end{table}

\begin{figure}[htbp]
  \centering
  \begin{minipage}[b]{0.48\textwidth}
    \centering
    \includegraphics[width=\textwidth]{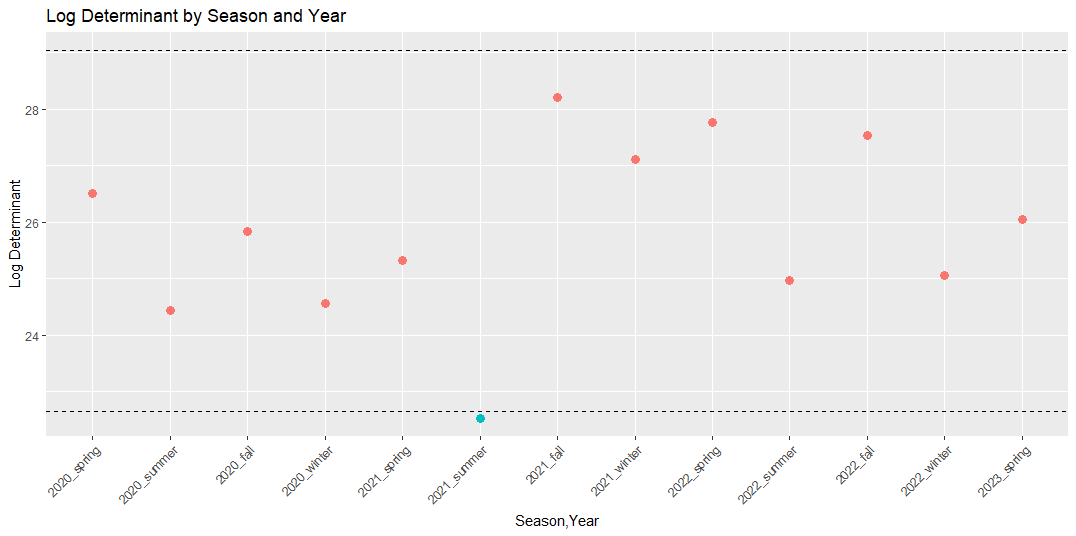}
    \label{fig:empirical covariance log determinant}
  \end{minipage}
  \hfill
  \begin{minipage}[b]{0.48\textwidth}
    \centering
    \includegraphics[width=\textwidth]{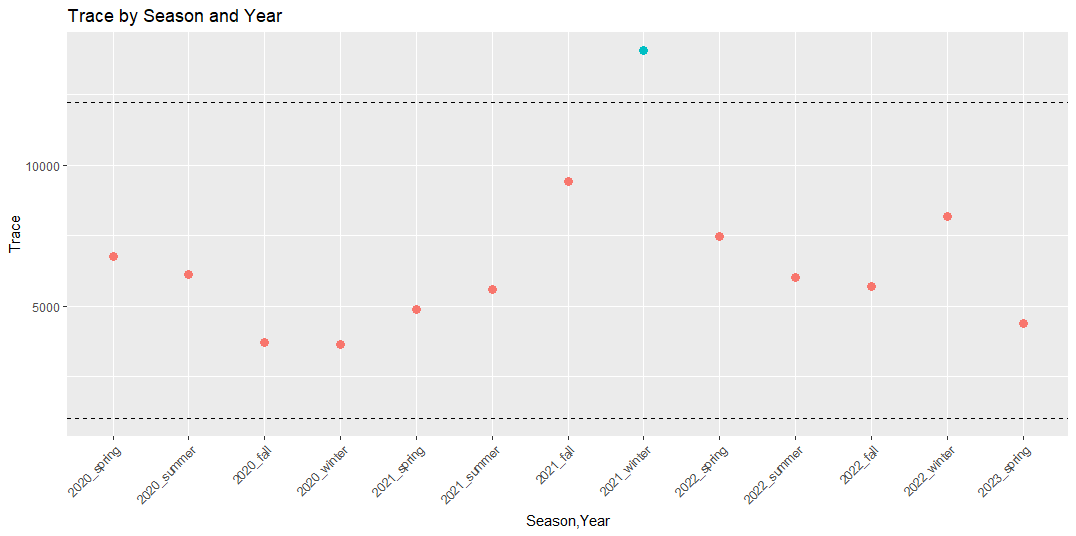}
    \label{fig:empirical covariance trace}
  \end{minipage}
  \caption{Log determinant (left) and trace (right) of $\hat{\Sigma}_{\gamma}$ by season and year. Black dotted lines highlight the interval $\mu(f(\hat{\Sigma}_{\gamma,s,y})) + 2 \sigma(f(\hat{\Sigma}_{\gamma,s,y}))$, where $f(\cdot)$ denotes correspondingly trace or log determinant. Points in red fall within the 2 sd interval, blue points are outliers for this region.}
  \label{fig: Prior validation}
\end{figure}

     As our analysis is focused specifically on the covariance of the coefficients for annual cyclic trends, we begin our analysis by evaluating the validity of our prior assumptions. In Figure (\ref{fig: Prior validation}), we fit the OLS model (\ref{eq: Independent OLS models}) across all season and year pairs, and compute $\mu(f(\hat{\Sigma}_{\gamma,s,y})) + 2\sigma(f(\hat{\Sigma}_{\gamma,s,y}))$, which are computed respectively across $s \in \{Spring,Summer, Fall, Winter\}$ and $y \in \{2020,2021,2022,2023\}$. Note that this prior validation loses some model interpretability of annual cyclic trends but provides valuable insights into the individual correlation patterns of the resulting sample covariance matrix of $\gamma$ across seasons. From the resulting covariance pattern, we can deduce that $tr(\Sigma_{\gamma,s,y})$ and $\log \vert \Sigma_{\gamma,s,y} \vert$ are approximately fixed over time.

     From the full OLS model in (\ref{eq: Independent OLS models}), we propose the following Bayesian joint model with a random effects coefficient for the annual cyclic patterns:
     \begin{align*} 
         \hat{Y}_{c,t,i} &= \alpha_{c,i}  + \sum_{j = 1}^{3} \big(\phi_{j,i} Y_{c,t-j,1} + \xi_{j,i} Y_{c,t - j, 2}\big) \\
         &+ \gamma_{1,i} \sin\big(2 \pi \frac{\Xi(t)}{365} \big) + \gamma_{2,i} \cos \big( 2 \pi \frac{\Xi (t)}{365} \big) + \epsilon_{c} \\
         \alpha_{c,i} &\sim N(0, \sigma_{\alpha})\\
         \phi_{j,1},\phi_{j,2} &\sim N(0, \sigma_{\phi}) \\
         \xi_{j,1},\xi_{j,2} &\sim N(0, \sigma_{\xi}) \\
         \gamma &\sim N(\mu, \Sigma)\\
         \mu &\sim N(0,\sigma_{\mu})\\
         \epsilon_{c} &\sim N^{+}(0,\sigma_{\epsilon}).
     \end{align*}

     In Table \ref{table: Residual Summary}, we give a comparative analysis of the residual to the mean for 3 models in the 48 cities:
     \begin{itemize}
         \item Model 1: {\tt auto.arima}, regressing $T_{MAX}$ and $T_{MIN}$ only on its own previous lags.
         \item Model 2: anisotropic diagonal covariance on $\gamma$ ($\Sigma = diag(\tau_{1}, \tau_{2}, \tau_{3}, \tau_{4}), \quad \tau_{\cdot}\stackrel{iid}{\sim} C^{+}(0,1)$).
         \item Model 3: A dynamic case where $\Sigma_{s}$ is parameterized according to $L_{t}^{\dagger}$ of Section \ref{sec: seasonally dynamic covariance} with $K = 4$, where two column stochastic matrix parameters $A_{1}$, $A_{2}$ are introduced to account for assumed regime shifts between Spring 2021 and Summer 2021 ($A_{1}, s \in \{5,6\}$) and a separate regime shift from Fall 2021 onward ($A_{2}, s \geq 7$).
     \end{itemize}

     For simplicity, we assume $\sigma_{\phi} = \sigma_{\xi} = 1$, $\sigma_{\alpha} = \sigma_{\mu} = \sigma_{\epsilon} = 5$ in each model. For each Bayesian model, the Bayesian R-squared (\cite{gelman2019r}) was not worse than the mean R-squared for independent OLS models.

    \begin{table}[ht]
  \centering
  \begin{tabular}{|c|c|c|c|c|c|c|}
    \hline
         Model & $\mu(r_{TMIN})$ & $\sigma(r_{TMIN})$ & $IQR(r_{TMIN})$ & $\mu(r_{TMAX})$ & $\sigma(r_{TMAX})$ & $IQR(r_{TMAX})$ \\
    \hline
    1 & .294 & 31.8 & 37.8 & .303 & 39.4 & 45.6 \\
    2 & -.0771 & 29.6 & 35.3 & .551 & 39.1 & 46.0 \\
    3 & -.0696 & 29.5 & 35.2 & .241 & 39.1 & 46.4  \\
    
    \hline
  \end{tabular}
  \caption{Summary Statistics of Residuals to Posterior Mean $\big(r = Y - (\mu \vert Y) \big)$. Columns are ordered across cities according to: mean residual, mean standard deviation of residual, mean IQR of residual for TMIN (first 3 columns) and TMAX (last 3 columns), respectively. Note that in all but $IQR(r_{TMAX})$, the seasonally dynamic model for $\gamma_{\cdot}$ outperformed all other models.}
  \label{table: Residual Summary}
\end{table}

\begin{figure}[h!]
        \centering
        \includegraphics[width=0.9\linewidth]{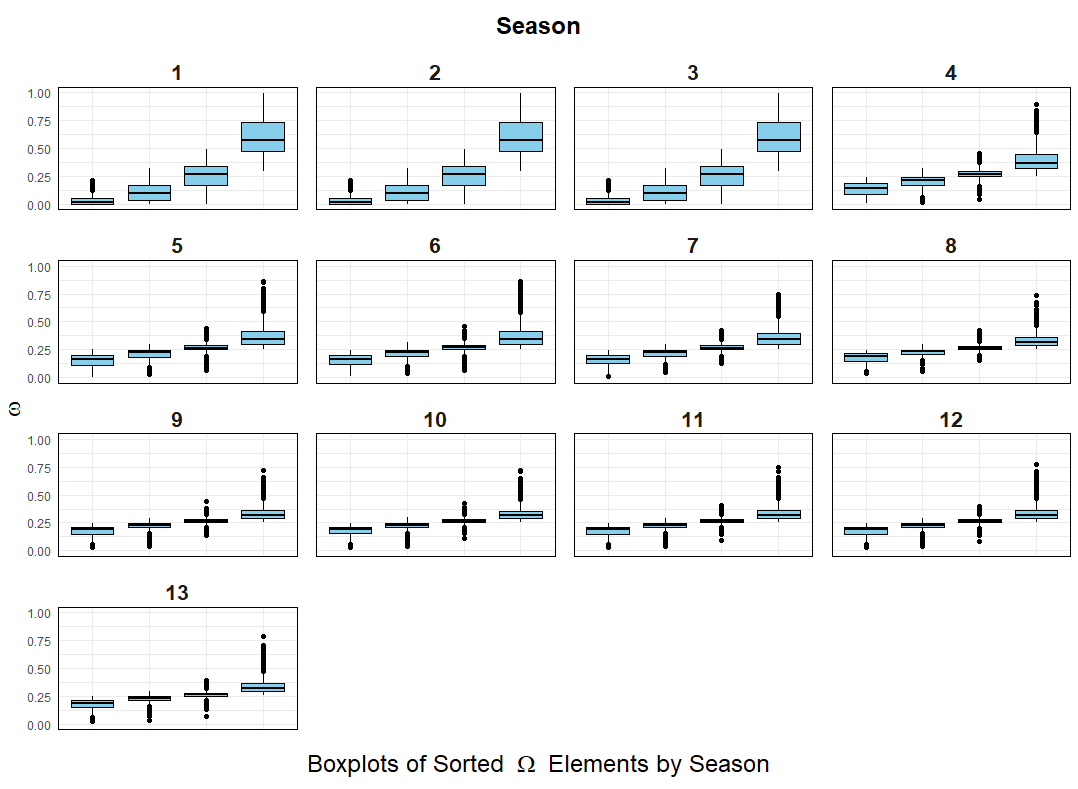}
        \caption{Sorted posterior $\Omega$ components under the SD-SCKPD model for $\gamma$ within our regression model. Note the stabilization of the regime shift at $\sim$ season 6, implicating a non-substantial regime shift for seasons $s \geq 7$. }
        \label{fig: Climate analysis omega}
    \end{figure}

\begin{figure}[h!]
  \centering
  \begin{minipage}{0.75\textwidth}
    \centering
    \includegraphics[width=\linewidth]{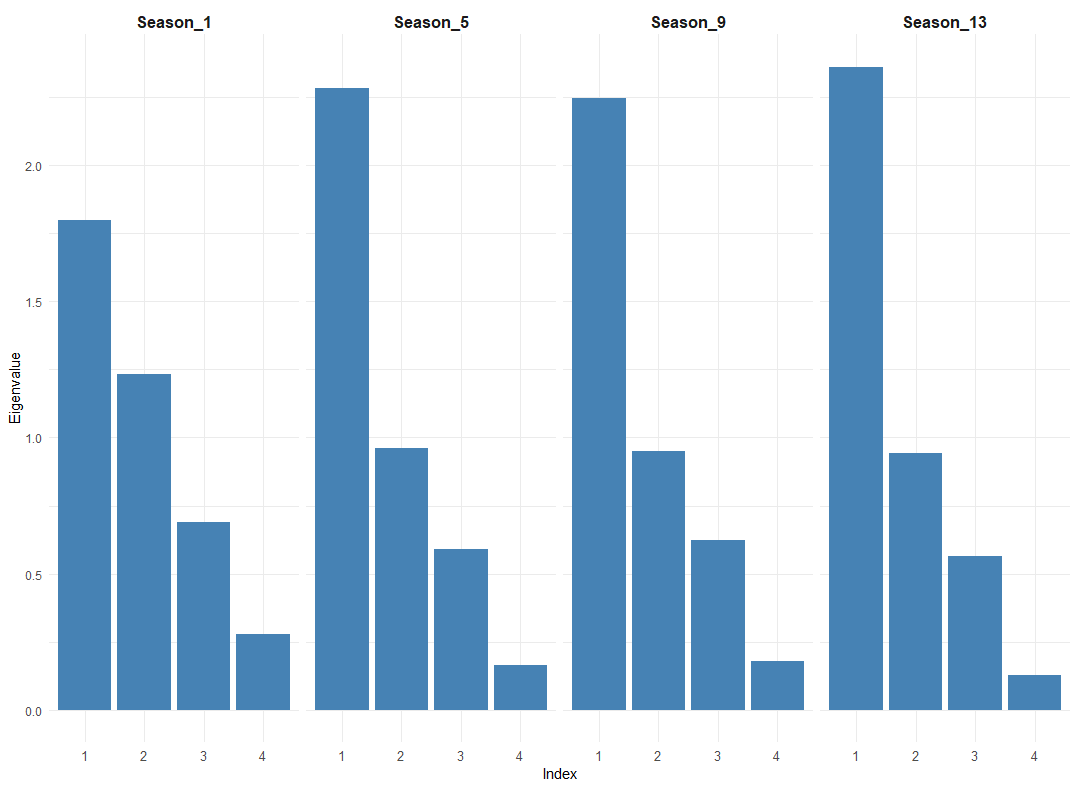}
    \vspace{0.5cm}
    \includegraphics[width=\linewidth]{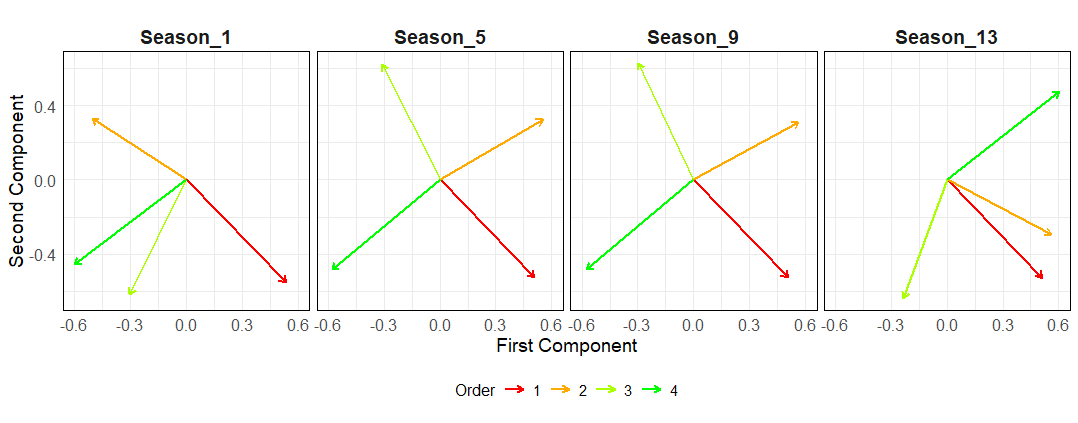}
  \end{minipage}
  \caption{Eigenvalues (top row) and eigenvectors (bottom row) of the resulting Correlation matrices $corr(L_{t} L_{t}^{T})$ for seasons 1,5,9, and 13. Performing a Procrustes analysis on the resulting eigenvectors between seasons 1 and all other seasons resulted in a maximal Procrustes sum of squares of $\approx 1^{-15}$ with a scaling of 1 and negligible translation. Hence, the eigenvectors themselves were all rotated versions of the resulting eigenvectors from season 1, but the changing eigenvalue structure then implies then implies non-negligible seasonal evolution of the correlation matrix with respect to the significance of principal directions.}
\end{figure}
    \section{Conclusion} \label{sec: conclusion}
    In this article, we explore the utility of a geometrically informed parameterization of Cholesky factors to relax the constraints imposed on the covariance of a tensor normal distribution. One potential avenue for future work is to consider a geodesic sampling approach. We noted that the posterior for this problem is indeed not geodesically convex. However, given the flexibility permitted for posterior distributions on the Cholesky factor, it may be interesting to consider such a sampling approach when the corresponding posterior is poorly conditioned. In such a case, it would be beneficial to consider further modifications, such as parallel tempering, to alleviate the nonconvexity of such an approach. If such an approach were considered, relatively few modifications would need to be made. In particular, the Lebesgue measure for posterior distributions defined in the Riemannian manifold is insufficient for posterior sampling \cite{byrne2013geodesic}. Instead, we need to resort to the Hausdorff measure according to the area formula \cite{federer2014geometric}:
    \[
    \mathcal{H}(dL\vert y) = \sqrt{\vert G(L) \vert} P(L \vert y) \lambda(dL)
    \]
    where $\lambda(\cdot)$ refers to the Lebesgue measure and $G(\cdot)$ is the matrix metric tensor. Note that in the product manifold geometry, $G(L)$ is particularly simple to compute in closed form. Assuming our position is specified in vector form as:
    \[
    vec(L) = (vec(\mathbb{D}(L)), vec(\lfloor L \rfloor))^{t}
    \]
    then for the norm
    \[
    \| V \|_{L} = \langle V,V\rangle_{F} + tr(\mathbb{D}(L)^{-1} \mathbb{D}(V) \mathbb{D}(L)^{-1} \mathbb{D}(V)) = V^{T} G(L) V.
    \]
    Gives the metric tensor in matrix form as:
    \[
    G(L) = \begin{pmatrix}
        \mathbb{D}(L)^{-1} \otimes \mathbb{D}(L)^{-1} & 0 \\
        0 & I_{\frac{d(d-1)}{2}}
    \end{pmatrix}
    \]
    which simply gives $\vert G(L) \vert = \vert \mathbb{D}(L) \vert^{2}$. Then the use of geodesic Monte Carlo algorithms a geodesic approach would be efficient in regards to only requiring inversion of diagonal matrices.

    Alternatively, exploring other metrics directly on SPD space could be worthwhile; the Cholesky parameterization explored here, while computationally tractable, yields cumbersome analytic gradients. With an SPD-valued metric, tractable gradients for the trace term were explored in \cite{simonis2025geodesicvariationalbayesmultiway}, which is parallelizable within a geodesic Monte Carlo implementation. We show that the log-Euclidean metric directly yields Fr\'echet means, which are multiway. Empirically, we found that the affine-invariant metric yields similar results. An alternative metric that we did not investigate was the Bures-Wasserstein metric \cite{bhatia2019bures}, which takes the form:
    \[
    g_{\Sigma}^{BW}(U,V) = \frac{1}{2} tr(\mathcal{L}_{\Sigma}[U] V)
    \]
    where $\mathcal{L}_{\Sigma}[U]$ denotes the Lyapunov operator, or the solution to the system of equations:
    \[
    \mathcal{L}_{\Sigma}[U]\Sigma + \Sigma \mathcal{L}_{\Sigma}[U] = U.
    \]
    For $A,B \in \mathcal{P}^{+}(d)$, the corresponding geodesic distance is given by:
    \[
    d^{BW}(A,B) = \big[ tr(A) + tr(B) -2tr(A^{\frac{1}{2}} B A^{\frac{1}{2}})^{\frac{1}{2}}]^{\frac{1}{2}}.
    \]
    
    To our knowledge, no closed form exists for the Fr\'echet mean of such a distance, and in such a case may be an interesting avenue for investigating differentiation directly through the $\arg\min$ operator, as was explored in \cite{lou2020differentiating}. However, we note that the matrix metric is given by \cite{han2021riemannian} as
    \[
    G_{BW}(\Sigma) = \Sigma \oplus \Sigma
    \]
    where $\oplus$ denotes the Kronecker sum operator:
    \[
    A \oplus B = A \otimes I + I \otimes B
    \]
    where an obvious difficulty with this then being the calculation of the corresponding Hausdorff measure due to the intractability of the determinant of the metric.

    A clear modification to the model assumption on $L^{\dagger}$ is to remove the assumed multiway structure on the diagonal. That is, instead model $L^{\dagger}$ as:
    \[
    L^{\dagger} = \sum_{i = 1}^{K} \lfloor L_{1}^{(i)}\otimes L_{2}^{(i)} \rfloor + \exp(D)
    \]
    for an unconstrained diagonal matrix $D$. While limiting in it's ability to detect true separability of a resulting covariance matrix, it provides an obvious generalization in flexibility at little increased computational complexity of the model. However, we note that many of the computational results regarding representation of the likelihood under the P-VL and the time-varying matrix normally distributed regression parameters techniques discussed in the supplement may require substantial modifications. 
    
    Lastly, this article has thus far made no mention of relationships to spatial approximations, where separability is a common assumption. When combined with the implications of how conditional independence in space-time models is well known to be driven by sparsity in the entries of Cholesky factors of precision or covariance matrices in such models through the Veccia approximation \cite{katzfuss2021general}, investigating methods for averaging over separable space-time functions could be a fruitful direction for investigation with respect to the scalability of the methods described in this article.
 
  %  \begin{acks}[Acknowledgments]
   %The authors would like to thank the anonymous referees, an Associate Editor and the Editor for their constructive comments that improved the quality of this paper.
% \end{acks}

\section{Supplementary Material} \label{sec: Computational results}
\subsection{Static Covariance} \label{subsec: Static covariance computational results} 
    In this section we consider several possible algebraic manipulations of the likelihood for efficient computation of the SCKPD Bayesian model. 

    We start by observing that if: 
    \begin{align*}
        \Sigma^{-1} &= LL^{T}\\
        L &= \sum_{i = 1}^{K} \lfloor L_{1}^{(i)} \otimes L_{2}^{(i)} \rfloor + D_{1} \otimes D_{2}.
    \end{align*}
    Then first note that 
    \begin{align*}
        LL^{T} &= \big[\sum_{k = 1}^{K} \lfloor L_{1}^{(k)} \otimes L_{2}^{(k)}\rfloor + D_{1} \otimes D_{2} \big]\big[\sum_{k = 1}^{K} \lfloor L_{1}^{(k)} \otimes L_{2}^{(k)}\rfloor + D_{1} \otimes D_{2} \big]^{T}\\
        &= \sum_{k_{1} = 1}^{k} \sum_{k_{2} = 1}^{k} \lfloor L_{1}^{(k_{1})} \otimes L_{2}^{(k_{1})}\rfloor \lfloor L_{1}^{(k_{2})} \otimes L_{2}^{(k_{2})}\rfloor^{T} \\
        &+ \sum_{k = 1}^{K} \lfloor L_{1}^{(k)} \otimes L_{2}^{(k)} \rfloor D_{1} \otimes D_{2} + \sum_{k = 1}^{K} \lfloor L_{1}^{(k)} \otimes L_{2}^{(k)} \rfloor^{T} D_{1} \otimes D_{2} \\
        &+ D_{1}^{2} \otimes D_{2}^{2}.
    \end{align*}
    Expanding the first term:
    \begin{align*}
        &\lfloor L_{1}^{(i)} \otimes L_{2}^{(i)}\rfloor \lfloor L_{1}^{(j)} \otimes L_{2}^{(j)}\rfloor\\ &= \big(\lfloor L_{1}^{(i)}\rfloor \otimes D_{2} + D_{1} \otimes \lfloor L_{2}^{(i)}\rfloor + D_{1} \otimes  D_{2}\big) \big(\lfloor L_{1}^{(j)} \rfloor^{T} \otimes D_{2} + D_{1} \otimes \lfloor L_{2}^{(j)} \rfloor^{T} + \lfloor L_{1}^{(j)}\rfloor^{T} \otimes \lfloor L_{2}^{(j)} \rfloor^{T}\big)\\
        &= \lfloor L_{1}^{(i)}\rfloor\lfloor L_{1}^{(j)}\rfloor^{T}\otimes D_{2}^{2} + \lfloor L_{1}^{(i)} \rfloor D_{1} \otimes D_{2} \lfloor L_{2}^{(j)}\rfloor^{T} + \lfloor L_{1}^{(i)} \rfloor\lfloor L_{1}^{(j)} \rfloor^{T} \otimes D_{2} \lfloor L_{2}^{(j)}\rfloor^{T} \\
        &+ D_{1}\lfloor L_{1}^{(j)} \rfloor^{T} \otimes \lfloor L_{2}^{(i)}\rfloor D_{2} + D_{1}^{2} \otimes \lfloor L_{2}^{(i)} \rfloor \lfloor L_{2}^{(j)} \rfloor^{T} + D_{1} \lfloor L_{1}^{(j)} \rfloor^{T} \otimes \lfloor L_{2}^{(i)} \rfloor \lfloor L_{2}^{(j)} \rfloor^{T}\\
        &+ D_{1} \lfloor L_{1}^{(j)}\rfloor^{T} \otimes D_{2}^{2} + D_{1}^{2} \otimes D_{2}\lfloor L_{2}^{(j)}\rfloor^{T} + D_{1}\lfloor L_{1}^{(j)}\rfloor^{T} \otimes D_{2} \lfloor L_{2}^{(j)} \rfloor^{T}.
    \end{align*}
    Likewise the second term is expanded as:
    \begin{align*}
        \lfloor L_{1}^{(k)} \otimes L_{2}^{(k)} \rfloor D_{1} \otimes D_{2} = \lfloor L_{1}^{(k)} \rfloor D_{1} \otimes D_{2}^{2} + D_{1}^{2} \otimes \lfloor L_{2}^{(k)} \rfloor D_{2} + \lfloor L_{1}^{(k)} \rfloor D_{1} \otimes \lfloor L_{2}^{(k)} \rfloor D_{2}
    \end{align*}
    and third term is given by:
    \begin{align*}
        \lfloor L_{1}^{(k)} \otimes L_{2}^{(k)} \rfloor^{T} D_{1} \otimes D_{2} = \lfloor L_{1}^{(k)} \rfloor^{T} D_{1} \otimes D_{2}^{2} + D_{1}^{2} \otimes \lfloor L_{2}^{(k)} \rfloor^{T} D_{2} + \lfloor L_{1}^{(k)} \rfloor^{T} D_{1} \otimes \lfloor L_{2}^{(k)} \rfloor^{T} D_{2}
    \end{align*}
    \begin{align*}
        tr(\Sigma^{-1} \sum_{i = 1}^{n} y_{i}y_{i}^{t}) = tr(\big[\sum_{k = 1}^{K} \lfloor L_{1}^{(k)} \otimes L_{2}^{(k)}\rfloor + D_{1} \otimes D_{2} \big]\big[\sum_{k = 1}^{K} \lfloor L_{1}^{(k)} \otimes L_{2}^{(k)}\rfloor + D_{1} \otimes D_{2} \big]^{T} \sum_{i = 1}^{n} y_{i}y_{i}^{T}).
    \end{align*}

    It is clear that these computations quickly become computationally cumbersome for finding an analytic solution, where in total we will produce $K^{2} + 2k + 1$ total summands, along with needing to construct $T^{(i)}(S, \cdot)$ for each term. As in the case of $D = 2$, we will instead take advantage of the P-VL decomposition in the same way as \cite{simonis2025separablegeodesiclagrangianmonte} to create a more compact form of the likelihood. Letting $\sum_{i = 1}^{N} y_{i} y_{i}^{T} = \sum_{q = 1}^{r^{2}} A_{q} \otimes B_{q}$, with $r = \min\{d_{1}, d_{2}\}$, direct calculation gives
    \begin{align*}
        L^{T} \sum_{i = 1}^{N} y_{i} y_{i}^{T}
        &= \sum_{q = 1}^{r^{2}} D_{1} A_{q} \otimes D_{2} B_{q} \\
        &+ \big(\sum_{ j = 1}^{K}  \lfloor L_{1}^{(j)} \rfloor^{T}A_{q} \otimes D_{2}B_{q} + D_{1} A_{q} \otimes \lfloor L_{2}^{(j)}\rfloor^{T} B_{q} + \lfloor L_{1}^{(j)} \rfloor^{T}A_{q} \otimes \lfloor L_{2}^{(j)} \rfloor^{T} B_{q}\big).
    \end{align*}
    From this, it is clear that
    \begin{align*}
        LL^{T} \sum_{i = 1}^{N} y_{i} y_{i}^{T} &= \sum_{q = 1}^{r^{2}} \big[D_{1}^{2} A_{q} \otimes D_{2}^{2} B_{q} \\
        &+ \big( \sum_{i = 1}^{K} \lfloor L_{1}^{(i)} \rfloor D_{1} A_{q} \otimes D_{2}^{2} B_{q}  + D_{1}^{2} A_{q} \otimes \lfloor L_{2}^{(i)} \rfloor D_{2} B_{q} + \lfloor L_{1}^{(i)} \rfloor D_{1} A_{q} \otimes \lfloor L_{2}^{(i)} \rfloor D_{2} B_{q} \big)\\
        &+ \big( \sum_{j = 1}^{K} D_{1}  \lfloor L_{1}^{(j)} \rfloor^{T} A_{q} \otimes D_{2}^{2} B_{q}    + D_{1}^{2} A_{q} \otimes  D_{2} \lfloor L_{2}^{(j)} \rfloor^{T} B_{q} +  D_{1} \lfloor L_{1}^{(j)} \rfloor^{T} A_{q} \otimes  D_{2} \lfloor L_{2}^{(j)} \rfloor^{T} B_{q} \big) \\
        &+ \sum_{i = 1}^{K} \sum_{j = 1}^{K} \big( \lfloor L_{1}^{(i)} \rfloor \lfloor L_{1}^{(j)} \rfloor^{T} A_{q} \otimes D_{2}^{2} B_{q} + \lfloor L_{1}^{(i)} \rfloor D_{1} A_{q} \otimes D_{2} \lfloor L_{2}^{(j)} \rfloor^{T} B_{q} \\
        &+ \lfloor L_{1}^{(i)}\rfloor \lfloor L_{1}^{(j)} \rfloor^{T} A_{q} \otimes \lfloor L_{2}^{(i)} \rfloor \lfloor L_{2}^{(j)} \rfloor^{T} B_{q} + D_{1} \lfloor L_{1}^{(j)} \rfloor^{T}A_{q} \otimes \lfloor L_{2}^{(i)} \rfloor D_{2} B_{q} \\
        &+ D_{1}^{2} A_{q} \otimes \lfloor L_{2}^{(i)} \rfloor \lfloor L_{2}^{(j)}\rfloor^{T} B_{q} + D_{1} \lfloor L_{1}^{(j)}\rfloor^{T} A_{q} \otimes \lfloor L_{2}^{(i)} \rfloor \lfloor L_{2}^{(j)}\rfloor^{T} B_{q} \\
        &+ \lfloor L_{1}^{(i)} \rfloor \lfloor L_{1}^{(j)} \rfloor^{T} A_{q} \otimes \lfloor L_{2}^{(i)} \rfloor D_{2} B_{q} + \lfloor L_{1}^{(i)} \rfloor D_{1} A_{q} \otimes \lfloor L_{2}^{(i)} \rfloor \lfloor L_{2}^{(j)} \rfloor^{T} B_{q}\\
        &+ \lfloor L_{1}^{(i)} \rfloor \lfloor L_{1}^{(j)}\rfloor^{T} A_{q} \otimes \lfloor L_{2}^{(i)} \rfloor \lfloor L_{2}^{(j)} \rfloor^{T} B_{q} \big) \big].\\
    \end{align*}
    
    By linearity of the trace and leveraging two properties of Kronecker products:
    \begin{align*}
        (A \otimes B)(C \otimes D) &= AC \otimes BD \\
        tr(A \otimes B) &= tr(A) tr(B)
    \end{align*}
    the trace is then expressible as:
    \begin{align*}
        tr(LL^{T} \sum_{i = 1}^{N} y_{i} y_{i}^{T}) &= \sum_{q = 1}^{r^{2}} \big[tr(D_{1}^{2} A_{q}) tr( D_{2}^{2} B_{q}) \\
        &+ \big( \sum_{i = 1}^{K} tr(\lfloor L_{1}^{(i)} \rfloor D_{1} A_{q}) tr( D_{2}^{2} B_{q})  + tr(D_{1}^{2} A_{q}) tr(\lfloor L_{2}^{(i)} \rfloor D_{2} B_{q}) \big)\\
        &+ \big( \sum_{i = 1}^{K}  tr(\lfloor L_{1}^{(i)} \rfloor D_{1} A_{q}) tr( \lfloor L_{2}^{(i)} \rfloor D_{2} B_{q}) \big)\\
        &+ \big(\sum_{j = 1}^{K} tr( D_{1}  \lfloor L_{1}^{(j)} \rfloor^{T} A_{q}) tr( D_{2}^{2} B_{q})    + tr(D_{1}^{2} A_{q}) tr(D_{2} \lfloor L_{2}^{(j)} \rfloor^{T} B_{q}) \\
        &+  tr(D_{1} \lfloor L_{1}^{(j)} \rfloor^{T} A_{q} ) tr(  D_{2} \lfloor L_{2}^{(j)} \rfloor^{T} B_{q}) \big) \\
        &+ \sum_{i = 1}^{K} \sum_{j = 1}^{K} \big( tr(\lfloor L_{1}^{(i)} \rfloor \lfloor L_{1}^{(j)} \rfloor^{T} A_{q}) tr( D_{2}^{2} B_{q}) + tr(\lfloor L_{1}^{(i)} \rfloor D_{1} A_{q}) tr( D_{2} \lfloor L_{2}^{(j)} \rfloor^{T} B_{q}) \\
        &+ tr(\lfloor L_{1}^{(i)}\rfloor \lfloor L_{1}^{(j)} \rfloor^{T} A_{q}) tr(\lfloor L_{2}^{(i)} \rfloor \lfloor L_{2}^{(j)} \rfloor^{T} B_{q}) + tr(D_{1} \lfloor L_{1}^{(j)} \rfloor^{T}A_{q}) tr( \lfloor L_{2}^{(i)} \rfloor D_{2} B_{q}) \\
        &+ tr(D_{1}^{2} A_{q})  tr(\lfloor L_{2}^{(i)} \rfloor \lfloor L_{2}^{(j)}\rfloor^{T} B_{q}) + tr(D_{1} \lfloor L_{1}^{(j)}\rfloor^{T} A_{q})tr( \lfloor L_{2}^{(i)} \rfloor \lfloor L_{2}^{(j)}\rfloor^{T} B_{q}) \\
        &+ tr(\lfloor L_{1}^{(i)} \rfloor \lfloor L_{1}^{(j)} \rfloor^{T} A_{q}) tr( \lfloor L_{2}^{(i)} \rfloor D_{2} B_{q}) + tr(\lfloor L_{1}^{(i)} \rfloor D_{1} A_{q}) tr( \lfloor L_{2}^{(i)} \rfloor \lfloor L_{2}^{(j)} \rfloor^{T} B_{q}) \\
        &+ tr(\lfloor L_{1}^{(i)} \rfloor \lfloor L_{1}^{(j)}\rfloor^{T} A_{q}) tr( \lfloor L_{2}^{(i)} \rfloor \lfloor L_{2}^{(j)} \rfloor^{T} B_{q}) \big) \big].\\
    \end{align*}
    While many summands are present within this, it's clear that each individual trace term is a sparse matrix multiplication. Moreover, we can leverage the fact that the P-VL decomposition inherits the following symmetry property
    \[
    S = \sum_{i = 1}^{r^{2}} A_{i} \otimes B_{i}, \quad S \in \mathcal{S}(d) \implies A_{i}, B_{i} \in \mathcal{S}(d_{1}) \times \mathcal{S}(d_{2}) \text{ for all } i.
    \]
    Using the above property, and noting $(ABC)^{T} = C^{T} B^{T} A^{T}$, then observe
    \[
    (\lfloor L_{t}^{(i)} \rfloor D_{t} C_{q})^{T} = C_{q} D_{t} \lfloor L_{t}^{(i)} \rfloor^{T}. 
    \]
    Further, by invariance of trace under cyclic permutations and transpositions, it is evident that \[
    tr(\lfloor L_{t}^{(i)} \rfloor D_{t} C_{q}) = tr(C_{q} D_{t} \lfloor L_{t}^{(i)} \rfloor^{T} ) = tr(D_{t} \lfloor L_{t}^{(i)}\rfloor^{T} C_{q}).
    \]
    
    Which gives an immediate simplification as:
    \begin{align*}
        &\big( \sum_{i = 1}^{K} tr(\lfloor L_{1}^{(i)} \rfloor D_{1} A_{q}) tr( D_{2}^{2} B_{q})  + tr(D_{1}^{2} A_{q}) tr(\lfloor L_{2}^{(i)} \rfloor D_{2} B_{q}) + tr(\lfloor L_{1}^{(i)} \rfloor D_{1} A_{q}) tr( \lfloor L_{2}^{(i)} \rfloor D_{2} B_{q}) \big)\\
        &+ \big(\sum_{j = 1}^{K} tr( D_{1}  \lfloor L_{1}^{(j)} \rfloor^{T} A_{q}) tr( D_{2}^{2} B_{q})    + tr(D_{1}^{2} A_{q}) tr(D_{2} \lfloor L_{2}^{(j)} \rfloor^{T} B_{q}) \\
        &+  tr(D_{1} \lfloor L_{1}^{(j)} \rfloor^{T} A_{q} ) tr(  D_{2} \lfloor L_{2}^{(j)} \rfloor^{T} B_{q}) \big) \\
        &= 2\big( \sum_{i = 1}^{K} tr(\lfloor L_{1}^{(i)} \rfloor D_{1} A_{q}) tr( D_{2}^{2} B_{q})  + tr(D_{1}^{2} A_{q}) tr(\lfloor L_{2}^{(i)} \rfloor D_{2} B_{q}) + tr(\lfloor L_{1}^{(i)} \rfloor D_{1} A_{q}) tr( \lfloor L_{2}^{(i)} \rfloor D_{2} B_{q}) \big).
    \end{align*}
    
    Now note that for the trace of a matrix product, we can write it in an order of complexity faster by leveraging the Hadamard product form
    \[
    tr(A B) = \sum_{i} \sum_{j} (A^{T} \odot B)_{i,j} = \sum_{i} \sum_{j} A[j,i] B[i,j].
    \]
    We will use $\sigma(A,B) = \sum_{i} \sum_{j} (A^{T} \odot B)_{i,j}$ to denote this less computationally taxing form. Then we can rewrite $tr(LL^{T} \sum_{i = 1}^{N} y_{i} y_{i}^{T})$ as:
    \begin{align*}
        tr(LL^{T} \sum_{i = 1}^{N} y_{i} y_{i}^{T}) &= \sum_{q = 1}^{r^{2}}\big[ \sigma(D_{1}^{2}, A_{q})\sigma(D_{2}^{2}, B_{q}) + \\
        &+ 2\sum_{i = 1}^{K} \sigma(\lfloor L_{1}^{(i)} \rfloor, D_{1} A_{q})\sigma(D_{2}^{2}, B_{q}) + \sigma(D_{1}^{2}, A_{q}) \sigma(\lfloor L_{2}^{(i)} \rfloor, D_{2} B_{q})  \\
        &+ 2\sum_{i = 1}^{K}  \sigma(\lfloor L_{1}^{(i)} \rfloor, D_{1} A_{q})\sigma(\lfloor L_{2}^{(i)} \rfloor, D_{2} B_{q}) \\
        &+ \sum_{i = 1}^{K} \sum_{j = 1}^{K} \big( \sigma(\lfloor L_{1}^{(i)} \rfloor, \lfloor L_{1}^{(j)} \rfloor^{T} A_{q})\sigma(D_{2}^{2}, B_{q}) + \sigma(\lfloor L_{1}^{(i)}\rfloor, D_{1} A_{q})\sigma(D_{2}, \lfloor L_{2}^{(j)} \rfloor^{T} B_{q}) \\
        &+ \sigma(\lfloor L_{1}^{(i)} \rfloor, \lfloor L_{1}^{(j)} \rfloor^{T} A_{q})\sigma(\lfloor L_{2}^{(i)} \rfloor, \lfloor L_{2}^{(j)}\rfloor^{T} B_{q}) + \sigma(D_{1}, \lfloor L_{1}^{(j)} \rfloor^{T} A_{q}) \sigma(\lfloor L_{2}^{(i)}\rfloor, \lfloor L_{2}^{(j)} \rfloor^{T} B_{q}) \\
        &+ \sigma(D_{1}^{2}, A_{q}) \sigma(\lfloor L_{2}^{(i)} \rfloor, \lfloor L_{2}^{(j)} \rfloor^{T} B_{q}) + \sigma(D_{1}, \lfloor L_{1}^{(j)} \rfloor^{T} A_{q})\sigma(\lfloor L_{2}^{(i)}\rfloor, \lfloor L_{2}^{(j)} \rfloor^{T} B_{q}) \\
        &+ \sigma(\lfloor L_{1}^{(i)} \rfloor, \lfloor L_{1}^{(j)} \rfloor^{T} A_{q}) \sigma(\lfloor L_{2}^{(i)} \rfloor, D_{2} B_{q}) + \sigma(\lfloor L_{1}^{(i)} \rfloor, D_{1} A_{q}) \sigma(\lfloor L_{2}^{(i)} \rfloor, \lfloor L_{2}^{(j)} \rfloor^{T} B_{q}) \\
        &+ \sigma(\lfloor L_{1}^{(i)} \rfloor, \lfloor L_{1}^{(j)}\rfloor^{T} A_{q}) \sigma(\lfloor L_{2}^{(i)} \rfloor, \lfloor L_{2}^{(j)} \rfloor^{T} B_{q}) \big)
        \big].
    \end{align*}
    Observe that each summand is then a Hadamard product between a sparse matrix and a matrix resulting from a sparse multiplication.

     Note that for the seasonally dynamic covariance model of Section \ref{sec: seasonally dynamic covariance}, no significant modifications need be made to the results from this section beyond replacing $\{L_{1}^{(i)}, L_{2}^{(i)}\}_{i = 1}^{K}$ with it's seasonal/cyclic counterparts $\{L_{1,(c,s)}^{(i)}, L_{2,(c,s)}^{(i)}\}$ which respectively act on $\{A_{q}^{(c,s)},B_{q}^{(c,s)}\}_{q = 1}^{r^{2}}$. Note this latter point simply involves for a given collection of observations $\{y_{1}^{(c,s)}, \ldots, y_{N}^{(c,s)}\}$ belonging to season $s$, cycle $c$ and performing the decomposition 
     \[
     \sum_{i = 1}^{N} y_{i}^{(c,s)} (y_{i}^{(c,s)})^{T} = \sum_{i = 1}^{r^{2}} A_{i}^{(c,s)} \otimes B_{i}^{(c,s)}
     \]
    \subsection{Seasonally Varying Parameterization} \label{subsec: computational details seasonally dynamic covariance}
    In this section we deal with the situation of a time varying mean. In particular, we assume the case of Section \ref{sec: Climate Analysis}, where 
    \begin{align}
        Y_{t} &\sim N(\beta X, \sigma^{2} I) \\
        \beta_{S} &\sim N((\bar{\beta}_{-S},\bar{\beta}_{S}), \Sigma)  
    \end{align}
    for some subset $\beta_{S} \subseteq \beta$, and
    \[ \Sigma = 
    \begin{pmatrix}
    \tau^{2} I_d & \boldsymbol{0} \\
    \boldsymbol{0} & L^{\dagger} {L^{\dagger}}^{T}.
    \end{pmatrix}
    \]
    Note that this can be a particularly efficient form compared to the data covariance case. Reparameterization trick can be used in two ways for this form. The less efficient but more direct way is through matrix-vector multiplication as:
    \[
    vec(\beta) \sim  (\tau x_{1}, L^{\dagger} x_{2}) + vec(\bar{\beta}) 
    \]
    where $x_{1},x_{2}$ are independently distributed as $x_{1}, x_{2} \sim N(0,I)$. However, note that the inefficiency of this is then attributed to the necessity of explicit computation of the Kronecker product for the strictly lower triangular entries for each $i \in \{1,\ldots, K\}$. The efficient representation of distributed computing from the previous section is infeasible, as it would require the iterative P-VL decomposition of the matricization of $x_{2}$ at each iteration.

    Instead, a more appealing form would come from the matrix normal form. This can be viewed directly from the matricization and vectorization:

    \begin{definition} \label{def: Matricization}
        Let $v \in \mathbb{R}^{d_{1} d_{2}}$, and define the {\bf mode-$\mathbf{i}$ folding} of $v$, $\mathcal{F}_{(i)}(v) = V_{(i)} \in \mathbb{R}^{d_{i} \times d_{-i}}$ element-wise as:
        \[
        [V_{(i)}]_{p,q} = v_{(p - 1)d_{i} + q}
        \]
        Define the {\bf mode-i unfolding} as the corresponding reverse mapping $\mathcal{UF}_{(i)}(V_{(i)}) \rightarrow v$
    \end{definition}
    
    From these definitions, in the case $D= 2$, for $x \in \mathbb{R}^{d_{1}d_{2}}$ such that $x  \sim N(0,I)$, then $ X_{(1)} = \mathcal{F}_{(1)}(x) \sim MN(0, I,I)$. Observing that the corresponding folding operation is a linear operator, it can easily be seen that:
    \begin{align*}
        L^{\dagger} x &= \big[ \sum_{i = 1}^{K} \mathbb{D}(L_{1}^{(i)})\otimes \lfloor L_{2}^{(i)} \rfloor + \lfloor L_{1}^{(i)} \rfloor \otimes \mathbb{D}(L_{2}) + \lfloor L_{1}^{(i)} \rfloor \otimes \lfloor L_{2}^{(i)} \rfloor + D_{1} \otimes D_{2}] x \\
        \implies \mathcal{F}_{(1)}(L^{\dagger}x) &= D_{2} F_{(1)}(x) D_{1} \\
        &+ \sum_{i = 1}^{K} \lfloor L_{2}^{(i)} \rfloor X_{(1)} D_{1} + D_{2} X_{(1)} \lfloor L_{1}^{(i)} \rfloor + \lfloor L_{2}^{(i)} \rfloor X_{(1)} \lfloor L_{1}^{(i)} \rfloor \\[12pt]
        \implies L^{\dagger} x &= \mathcal{UF}_{(1)}(\mathcal{F}_{(1)}(L^{\dagger} x)) = \mathcal{UF}_{(1)}(D_{2} X_{(1)} D_{1}) \\
        &+ \sum_{i = 1}^{K} \mathcal{UF}_{(1)}(\lfloor L_{2}^{(i)} \rfloor X_{(1)} D_{1}) + \mathcal{UF}_{(1)}(D_{2} X_{(1)} \lfloor L_{1}^{(i)} \rfloor) + \mathcal{UF}_{(1)}(\lfloor L_{2}^{(i)} \rfloor X_{(1)} \lfloor L_{1}^{(i)} \rfloor ).
    \end{align*} 
    Note this form is much more efficient than through the use of P-VL, as gradient complexity no longer scales multiplicatively with $n > 1$ observations.

%%%%%%%%%%%%%%%%%%%%%%%%%%%%%%%%%%%%%%%%%%%%%%
%% Funding information, if any,             %%
%% should be provided in the                %%
%% funding section.                         %%
%%%%%%%%%%%%%%%%%%%%%%%%%%%%%%%%%%%%%%%%%%%%%%
\textbf{Funding:}
This research was partially supported by a grant from New Presbyterian Hospital and the National Institute of Allergy and Infectious Diseases of the National Institutes of Health under award number P01AI159402.

%%%%%%%%%%%%%%%%%%%%%%%%%%%%%%%%%%%%%%%%%%%%%%
%% Supplementary Material, including data   %%
%% sets and code, should be provided in     %%
%% {supplement} environment with title      %%
%% and short description. It cannot be      %%
%% available exclusively as external link.  %%
%% All Supplementary Material must be       %%
%% available to the reader on Project       %%
%% Euclid with the published article.       %%
%%%%%%%%%%%%%%%%%%%%%%%%%%%%%%%%%%%%%%%%%%%%%%

\bibliographystyle{plain}
\bibliography{sample-base}

\begin{thebibliography}{10}

\bibitem{agarap2018breast}
Abien Fred~M Agarap.
\newblock On breast cancer detection: an application of machine learning algorithms on the {W}isconsin diagnostic dataset.
\newblock In {\em Proceedings of the 2nd International Conference on Machine Learning and Soft Computing}, pages 5--9, 2018.

\bibitem{bhatia2019bures}
Rajendra Bhatia, Tanvi Jain, and Yongdo Lim.
\newblock On the bures--wasserstein distance between positive definite matrices.
\newblock {\em Expositiones Mathematicae}, 37(2):165--191, 2019.

\bibitem{brown2000blur}
Patrick~E Brown, Gareth~O Roberts, Kjetil~F K{\aa}resen, and Stefano Tonellato.
\newblock Blur-generated non-separable space--time models.
\newblock {\em Journal of the Royal Statistical Society Series B: Statistical Methodology}, 62(4):847--860, 2000.

\bibitem{byrne2013geodesic}
Simon Byrne and Mark Girolami.
\newblock Geodesic monte carlo on embedded manifolds.
\newblock {\em Scandinavian Journal of Statistics}, 40(4):825--845, 2013.

\bibitem{chen2021space}
Wanfang Chen, Marc~G Genton, and Ying Sun.
\newblock Space-time covariance structures and models.
\newblock {\em Annual Review of Statistics and Its Application}, 8(1):191--215, 2021.

\bibitem{devroye2006nonuniform}
Luc Devroye.
\newblock Nonuniform random variate generation.
\newblock {\em Handbooks in Operations Research and Management Science}, 13:83--121, 2006.

\bibitem{federer2014geometric}
Herbert Federer.
\newblock {\em Geometric Measure Theory}.
\newblock Springer, 2014.

\bibitem{fosdick2014separable}
Bailey~K Fosdick and Peter~D Hoff.
\newblock Separable factor analysis with applications to mortality data.
\newblock {\em The Annals of Applied Statistics}, 8(1):120, 2014.

\bibitem{gelfand2000gibbs}
Alan~E Gelfand.
\newblock Gibbs sampling.
\newblock {\em Journal of the American Statistical Association}, 95(452):1300--1304, 2000.

\bibitem{gelman2019r}
Andrew Gelman, Ben Goodrich, Jonah Gabry, and Aki Vehtari.
\newblock R-squared for {B}ayesian regression models.
\newblock {\em The American Statistician}, 2019.

\bibitem{genton2007separable}
Marc~G Genton.
\newblock Separable approximations of space-time covariance matrices.
\newblock {\em Environmetrics: The Official Journal of the International Environmetrics Society}, 18(7):681--695, 2007.

\bibitem{hall2013lie}
Brian~C Hall and Brian~C Hall.
\newblock {\em Lie Groups, Lie Algebras, and Representations}.
\newblock Springer, 2013.

\bibitem{han2021riemannian}
Andi Han, Bamdev Mishra, Pratik~Kumar Jawanpuria, and Junbin Gao.
\newblock On riemannian optimization over positive definite matrices with the bures-wasserstein geometry.
\newblock {\em Advances in Neural Information Processing Systems}, 34:8940--8953, 2021.

\bibitem{hasselman2018package}
Berend Hasselman and Maintainer~Berend Hasselman.
\newblock Package ‘nleqslv’.
\newblock {\em R package version}, 3(2), 2018.

\bibitem{hatfield2018separable}
Laura~A Hatfield and Alan~M Zaslavsky.
\newblock Separable covariance models for health care quality measures across years and topics.
\newblock {\em Statistics in Medicine}, 37(12):2053--2066, 2018.

\bibitem{hitchcock1927expression}
Frank~L Hitchcock.
\newblock The expression of a tensor or a polyadic as a sum of products.
\newblock {\em Journal of Mathematics and Physics}, 6(1-4):164--189, 1927.

\bibitem{hofer2012symplectic}
Helmut Hofer and Eduard Zehnder.
\newblock {\em Symplectic Invariants and Hamiltonian Dynamics}.
\newblock Birkh{\"a}user, 2012.

\bibitem{hoff2011separable}
Peter~D Hoff.
\newblock Separable covariance arrays via the tucker product, with applications to multivariate relational data.
\newblock {\em Bayesian Analysis}, 2011.

\bibitem{holbrook2018geodesic}
Andrew Holbrook, Shiwei Lan, Alexander Vandenberg-Rodes, and Babak Shahbaba.
\newblock Geodesic lagrangian monte carlo over the space of positive definite matrices: with application to {B}ayesian spectral density estimation.
\newblock {\em Journal of Statistical Computation and Simulation}, 88(5):982--1002, 2018.

\bibitem{hut1995building}
Piet Hut, Jun Makino, and Steve McMillan.
\newblock Building a better leapfrog.
\newblock {\em Astrophysical Journal, Part 2-Letters (ISSN 0004-637X), vol. 443, no. 2, p. L93-L96}, 443:L93--L96, 1995.

\bibitem{katzfuss2021general}
Matthias Katzfuss and Joseph Guinness.
\newblock A general framework for vecchia approximations of gaussian processes.
\newblock {\em Statistical Science}, 36(1):124--141, 2021.

\bibitem{kolda2006multilinear}
Tamara~Gibson Kolda.
\newblock Multilinear operators for higher-order decompositions.
\newblock Technical report, Sandia National Laboratories (SNL), Albuquerque, NM, and Livermore, CA, 2006.

\bibitem{lin2019riemannian}
Zhenhua Lin.
\newblock Riemannian geometry of symmetric positive definite matrices via cholesky decomposition.
\newblock {\em SIAM Journal on Matrix Analysis and Applications}, 40(4):1353--1370, 2019.

\bibitem{lou2020differentiating}
Aaron Lou, Isay Katsman, Qingxuan Jiang, Serge Belongie, Ser-Nam Lim, and Christopher De~Sa.
\newblock Differentiating through the fr{\'e}chet mean.
\newblock In {\em International Conference on Machine Learning}, pages 6393--6403. PMLR, 2020.

\bibitem{neal2011mcmc}
Radford~M Neal et~al.
\newblock Mcmc using hamiltonian dynamics.
\newblock {\em Handbook of Markov Chain Monte Carlo}, 2(11):2, 2011.

\bibitem{pitsanis1997kronecker}
N.~P. Pitsanis.
\newblock {\em The {K}ronecker product in approximation and fast transform generation}.
\newblock PhD thesis, Cornell University, Ithaca, NY, 1997.

\bibitem{simonis2025geodesicvariationalbayesmultiway}
Quinn Simonis and Martin~T. Wells.
\newblock Geodesic variational {B}ayes for multiway covariances, 2025.

\bibitem{simonis2025separablegeodesiclagrangianmonte}
Quinn Simonis and Martin~T. Wells.
\newblock Separable geodesic lagrangian monte carlo for inference in 2-way covariance models, 2025.

\bibitem{song2023separability}
Dogyoon Song and Alfred~O Hero.
\newblock On separability of covariance in multiway data analysis.
\newblock {\em arXiv preprint arXiv:2302.02415}, 2023.

\bibitem{tsiligkaridis2013covariance}
Theodoros Tsiligkaridis and Alfred~O Hero.
\newblock Covariance estimation in high dimensions via {K}ronecker product expansions.
\newblock {\em IEEE Transactions on Signal Processing}, 61(21):5347--5360, 2013.

\bibitem{tucker1966some}
Ledyard~R Tucker.
\newblock Some mathematical notes on three-mode factor analysis.
\newblock {\em Psychometrika}, 31(3):279--311, 1966.

\bibitem{utpala2022differentially}
Saiteja Utpala, Praneeth Vepakomma, and Nina Miolane.
\newblock Differentially private fr$\backslash$'echet mean on the manifold of symmetric positive definite (spd) matrices with log-euclidean metric.
\newblock {\em arXiv preprint arXiv:2208.04245}, 2022.

\bibitem{VanLoan1993}
C.~F. Van~Loan and N.~Pitsianis.
\newblock Approximation with {K}ronecker products.
\newblock In Marc~S. Moonen, Gene~H. Golub, and Bart L.~R. De~Moor, editors, {\em Linear Algebra for Large Scale and Real-Time Applications}, pages 293--314. Springer Netherlands, Dordrecht, 1993.

\bibitem{wiesel2012convexity}
Ami Wiesel.
\newblock On the convexity in {K}ronecker structured covariance estimation.
\newblock In {\em 2012 IEEE Statistical Signal Processing Workshop (SSP)}, pages 880--883. IEEE, 2012.

\end{thebibliography}

\end{document}